\documentclass[traditabstract]{aa}
\usepackage{natbib}
\usepackage{epsfig}
\usepackage{txfonts}
\usepackage{graphicx}
\usepackage{txfonts}

\usepackage{epsf}
\usepackage{rotating}
\usepackage{graphics}
\usepackage[final]{pdfpages}

\def\teff{\hbox{$T_{\rm eff}$}}
\def\logg{\hbox{$\log g$}}

\def\rhk{\hbox{$\log R'_{\mathrm{HK}}$}}
\def\ms{\,m\,s$^{-1}$}         %m.s -1
\def\ms{\hbox{\,m\,s$^{-1}$}}         %m.s -1
\def\m2s2{\hbox{\,m$^{2}$\,s$^{-2}$}} %m2.s -2
\def\Msun{\hbox{M$_{\odot}$}}             %Msun

\def\Mjup{\hbox{$\mathrm{M}_{\rm Jup}$}}
\def\Rjup{\hbox{$\mathrm{R}_{\rm Jup}$}}
\def\me{\hbox{$\mathrm{M}_{\oplus}$}}

\def \mp{$M_{\rm p}$}

\def\a{HD\,13908}
\def\b{HD\,159243}
\def \c{HIP\,91258}

\def \1s{$1\,\sigma$}

\def \t0{T$_0$}
\def \mp{$M_{\rm p}$}

\begin{document}

\title{The SOPHIE search for northern extrasolar planets \thanks{Based on observations collected with the SOPHIE spectrograph on the 1.93-m telescope at Observatoire de Haute-Provence (CNRS), France, by the SOPHIE RPE Consortium (program PNP.CONS). Table 8 is only available in electronic form at the CDS via anonymous ftp to cdsarc.u-strasbg.fr (130.79.128.5).}\\
VI. Three new hot Jupiters in multi-planet extrasolar systems}

\author{
Moutou,~C.\inst{1}
\and H\'ebrard,~G.\inst{2,3}
\and Bouchy,~F.\inst{1,4}
\and Arnold, L. \inst{3}
\and Santos,~N.C.\inst{5,10}
\and Astudillo-Defru,~N.\inst{6}
\and Boisse,~I.\inst{5}
\and Bonfils,~X.\inst{6}
\and Borgniet,~S.\inst{6}
\and Delfosse,~X.\inst{6}
\and D\'iaz, R. F.\inst{1}
\and Ehrenreich,~D.\inst{4}
\and Forveille,~T.\inst{6} 
\and Gregorio,~J.\inst{7}
\and Labrevoir,~O.\inst{8}
\and Lagrange,~A.-M.\inst{6}
%\and Lovis,~C.\inst{4}
\and Montagnier,~G.\inst{3,2}
\and Montalto,~M.\inst{5} 
\and Pepe,~F.\inst{4}
\and Sahlmann,~J.\inst{4}
\and Santerne, ~A.\inst{5}
\and S\'egransan,~D.\inst{4}
\and Udry,~S.\inst{4}
\and Vanhuysse,~M.\inst{9}
%\and Vidal-Madjar,~A.\inst{2} 
}
\offprints{Claire.Moutou@lam.fr}

\institute{
Aix Marseille University, CNRS, LAM (Laboratoire d'Astrophysique de Marseille) UMR 7326, 13388 Marseille cedex 13, France
\email{Claire.Moutou@oamp.fr}
\and Institut d'Astrophysique de Paris, UMR 7095 CNRS, Universit\'e Pierre \& Marie Curie, 98bis boulevard Arago, 75014 Paris, France
\and Observatoire de Haute-Provence, CNRS \& OAMP, 04870 Saint-Michel l'Observatoire, France
\and Observatoire de Gen\`eve, Universit\'e de Gen\`eve, 51 Chemin des Maillettes, 1290 Sauverny, Switzerland
\and Centro de Astrof\'isica, Universidade do Porto, Rua das Estrelas, 4150-762 Porto, Portugal
\and UJF-Grenoble 1 / CNRS-INSU, Institut de Plan\'etologie et d'Astrophysique de Grenoble (IPAG) UMR 5274, Grenoble, F-38041, France 
\and Atalaia group, Crow Observatory-Portalegre, Portugal
%\and Association Adagio, L'Observatoire, 31540 Belesta Lauragais, France
%\and Universit\'e Toulouse III, UMR5187, 118 route de Narbonne 31062 Toulouse, France
\and Centre d'Astronomie, Plateau du Moulin \`a Vent, 04870 Saint-Michel-l'Observatoire, France
\and Oversky, 47 All\'ee des Palanques, 33127 Saint Jean d'Illac, France
\and Departamento de F\'isica e Astronomia, Faculdade de Ci\^encias, Universidade do Porto, Portugal
}
%   \date{Received ; accepted }
 
  \abstract{  We present high-precision radial-velocity measurements of three solar-type stars: \a, \b, and \c. The observations were made with the SOPHIE spectrograph  at the 1.93-m telescope of Observatoire de Haute-Provence (France). They show that these three bright stars host exoplanetary systems composed of at least two companions. \a\,b is a planet with a minimum mass of 0.865$\pm$0.035 \Mjup\, on a circular orbit with a period of 19.382$\pm$0.006 days. There is an outer massive companion in the system with a period of 931$\pm$17 days, $e$ = 0.12$\pm$0.02, and a minimum mass of 5.13$\pm$0.25 \Mjup. The star \b\, also has two detected companions with respective masses, periods, and eccentricities of \mp = 1.13$\pm$0.05 and 1.9$\pm$0.13 \Mjup, $P$ = 12.620$\pm$0.004 and 248.4$\pm$4.9 days, and $e$ = 0.02$\pm$0.02 and 0.075$\pm$0.05. Finally, the star \c\, has a planetary companion with a minimum mass of 1.068$\pm$0.038 \Mjup, an orbital period of 5.0505$\pm$0.0015 days, and a quadratic trend indicating an outer planetary or stellar companion that is as yet uncharacterized. The planet-hosting stars \a, \b, and \c\  are main-sequence stars of spectral types F8V, G0V, and G5V, respectively, with moderate activity levels. \c\, is slightly over-metallic, while the two other stars have solar-like metallicity. The three systems are discussed in the frame of formation and dynamical evolution models of systems composed of several giant planets.
}
\titlerunning{Three new hot Jupiters in extrasolar systems}
\authorrunning{C. Moutou et al.}

\maketitle

\section{Introduction}
Radial-velocity surveys have continuously provided new extrasolar planetary candidates since 1995 \citep{mayor95}. The observing strategy, consisting of multiple observations of single stars, is time-consuming and requires long-term programs on quasi-dedicated telescopes. These observations are necessary, however,  to achieve a complete picture of the exoplanet population, especially towards the most populated long-period or/and low-mass ends. Statistical analyses were recently published by \citet{mayor}, \citet{howard}, and \citet{wright12}, indicating the frequency of stars that harbour close-in massive planets - also called hot Jupiters: \citet{mayor} reported a value of 0.89$\pm$0.36\% for the occurrence of planets more massive than 50\me\, that have periods shorter than 11 days, \citet{howard} found an occurrence of 1.2$\pm$0.2\% for planets more massive than 100\me\, with periods shorter than 12 days and \citet{wright12} estimated 1.2$\pm$0.38\% for planets more massive than 30\me\, with periods shorter than 10 days. In the Kepler survey, which has a different target sample, 0.43\% of stars are found to harbour giant planets with radii between 6 and 22 Earth radii and periods shorter than 10 days \citep{howard12,fressin13}. The frequency of planets then increases towards lower masses, to reach  more than 50\% of the stars \citep{mayor} within the current detection limits of the available instruments. Hot Jupiters are thus extremely rare objects, although they have been at the heart of exoplanetary science for more than fifteen years, particularly for detailed characterizations, such as atmospheric studies and observations. 

In this paper, we report the discovery of three new planets with orbital periods shorter than 20 days and projected masses of about one Jupiter mass. Interestingly, they are the inner planets in a multiple system, with a more massive second companion in the external part of the system. These discoveries were made in the context of a large program with SOPHIE, the high-precision radial-velocity spectrograph at the Observatoire de Haute-Provence. The survey for giant planets occupies about 12\% of the telescope time since 2006 and has already allowed the discovery of twelve companions in the planet or brown dwarf range \citep{dasilva07,santos08,bouchy09,hebrard10,boisse10,diaz12}. In Section \ref{spec}, we describe the observations and analyses, in Section\,Ê\ref{stars} we report the stellar properties, and in Section \ref{rv} we show and analyse the radial-velocity observations. In Section \ref{disc}, we discuss the results and conclude.

\section{Spectroscopic observations}
\label{spec}

The spectroscopic observations were obtained with the SOPHIE spectrograph  at the Observatoire de Haute-Provence with the 1.93-m telescope in the framework of the large program (PNP.CONS) led by the SOPHIE Exoplanet Consortium, presented in detail by \citet{bouchy09}. The SOPHIE instrument \citep{perruchot11} is a fiber-fed environmentally stabilized echelle spectrograph covering the visible range from 387 to 694 nm. The spectral resolving power is 75,000 in  high-resolution mode. 

The radial velocities (RV) are obtained by cross-correlating the extracted spectra with a numerical mask based on stellar spectra, following the method developed by \citet{baranne} and \citet{pepe02}. For all three stars the numerical mask corresponding to a G2 spectral type was used because it provides the smallest individual error bars and residual scatter compared with the other available F0 and K5 masks.

The spectrograph has been upgraded in two steps \citep{bouchy13}, resulting in a slight zero-point motion at dates JD-2,400,000 = 55,730 and 56,274. The first update corresponds to the installation of a section of octagonal fibers upstream of the double scrambler, the second one to another section of octagonal fibers downstream; SOPHIE+ labels the measurements obtained after the updates. When applicable, the data sets should thus be taken as independent, with a free offset to be adjusted together with the Keplerian solution.  The velocity offset between SOPHIE and SOPHIE+ is 11$\pm$10\ms for \a\ and 37$\pm$12m/s for \b. They agree at 1.2$\sigma$, converging into a 24$\pm$12\ms\ velocity shift for solar-like stars. The error on this offset propagates into the parameters of the orbits, with the largest impact on the errors on the mass and period of the outer planets. 
%This offset is of the order of 20 \ms. As its exact value may depend on the stellar spectrum, it is better to let it vary freely in the orbit fitting. Note that after this upgrade, the radial-velocity accuracy routinely reaches 2 \ms\,Êfor stars of magnitude less than 8.5. This is achieved using the Thorium-Argon simultaneous observations and an exposure time of 15 minutes, that allows damping the impact of short-term stellar oscillations. 

In the context of the program where \a, \b, and \c\ were observed, the objective was to obtain multiple measurements on a large sample of stars at a moderate RV precision. A signal-to-noise ratio of 50 only is aimed at, which gives an average noise level of 5 \ms\ for individual measurements. 

%{\bf With the observing template used for these observations, the sky background is simultaneously monitored in a neighbour fiber.} 
%It helps estimating the sky background contribution and, if necessary, correcting for it. In our data sets, only 10 to 30\% of the measurements 
%A  correction for this background is required when the sky brightness is high and the Earth velocity is not enough separated from the stellar velocity. In the measurements analyzed in this article, the exposure times are short (5 min average) due to the target brightness and goal signal-to-noise ratio, and the background level is much fainter than the stellar flux. The radial velocities are thus not significantly affected by the background light. Even though it is sometimes detected in the spectra, the level of calibrated correction would be of the order of a few \ms. This is not worth a correction but may contribute to the residual jitter.

The radial-velocity measurements of the three stars are given in Tables \ref{rva}, \ref{rvb} and \ref{rvc}.

\begin{figure}[h!]
\begin{center}
\centering
\epsfig{file=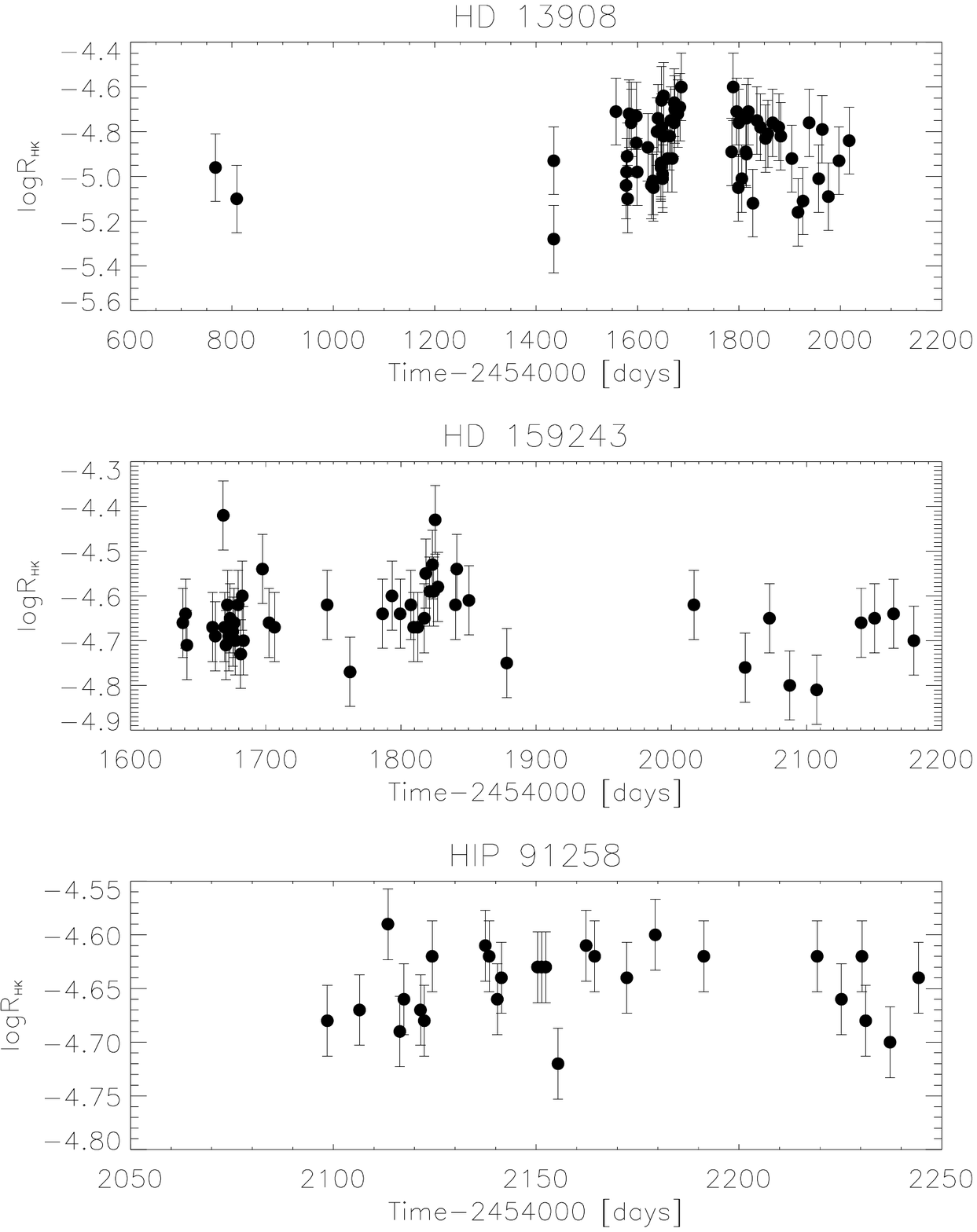,width=0.6\textwidth}
\vspace{-2.5cm}
\caption{Variations with time of the \rhk\ activity index for \a\ (top), \b\ (middle), and \c\ (bottom). }
\label{logr}
\end{center}
\end{figure}

\begin{figure}[h!]
\epsfig{file=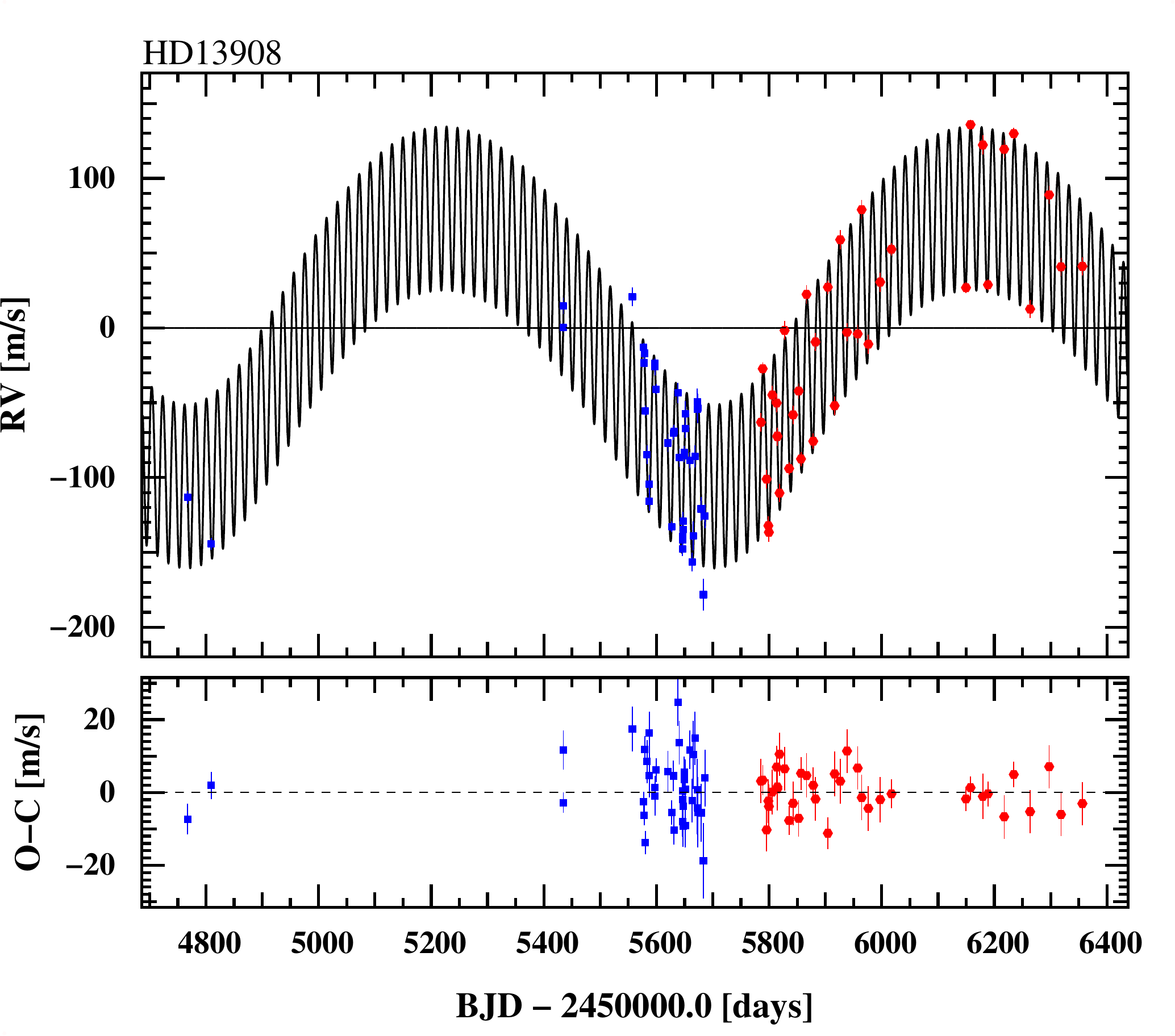,width=0.4\textwidth}
\epsfig{file=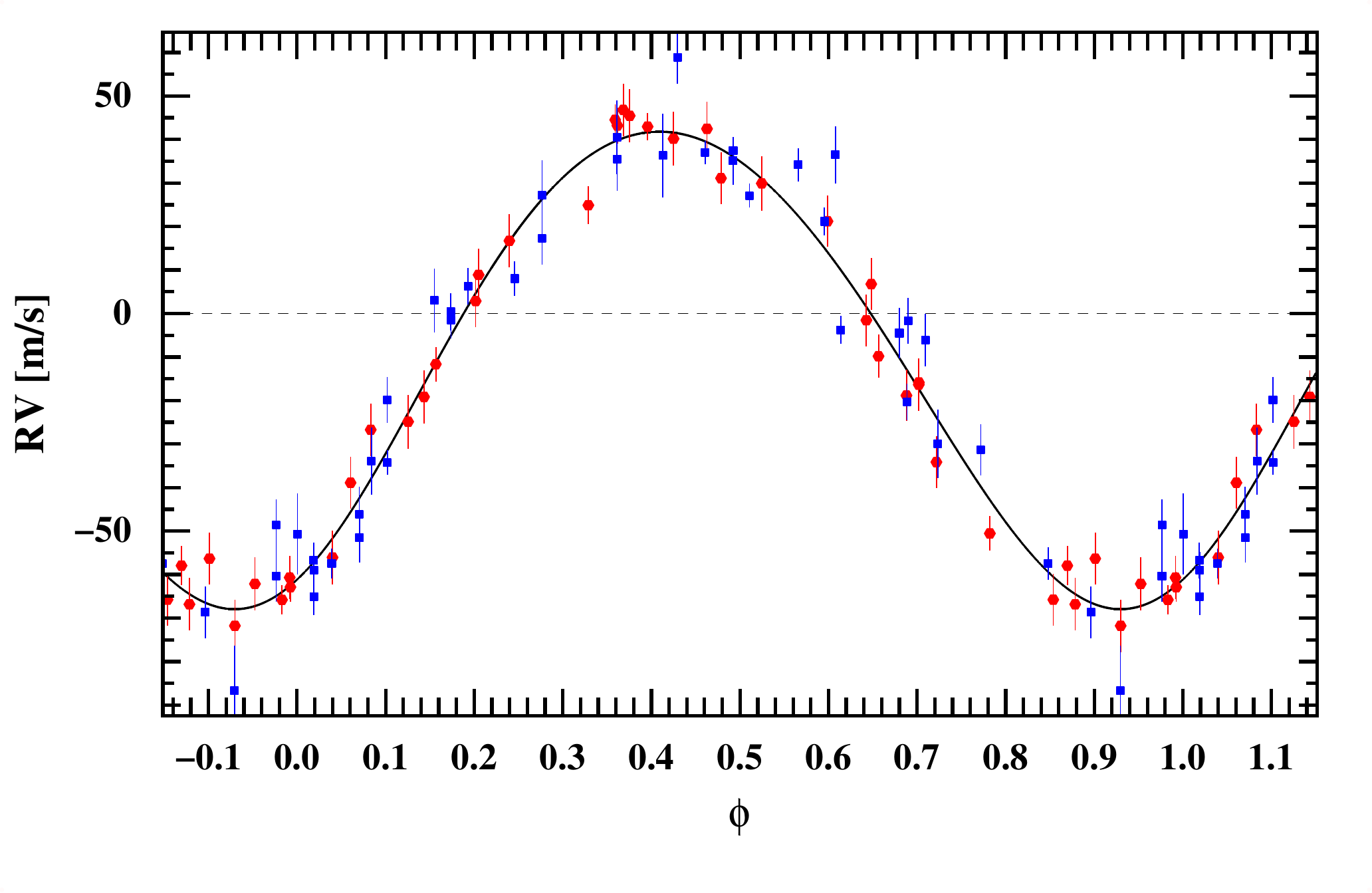,width=0.4\textwidth}
\epsfig{file=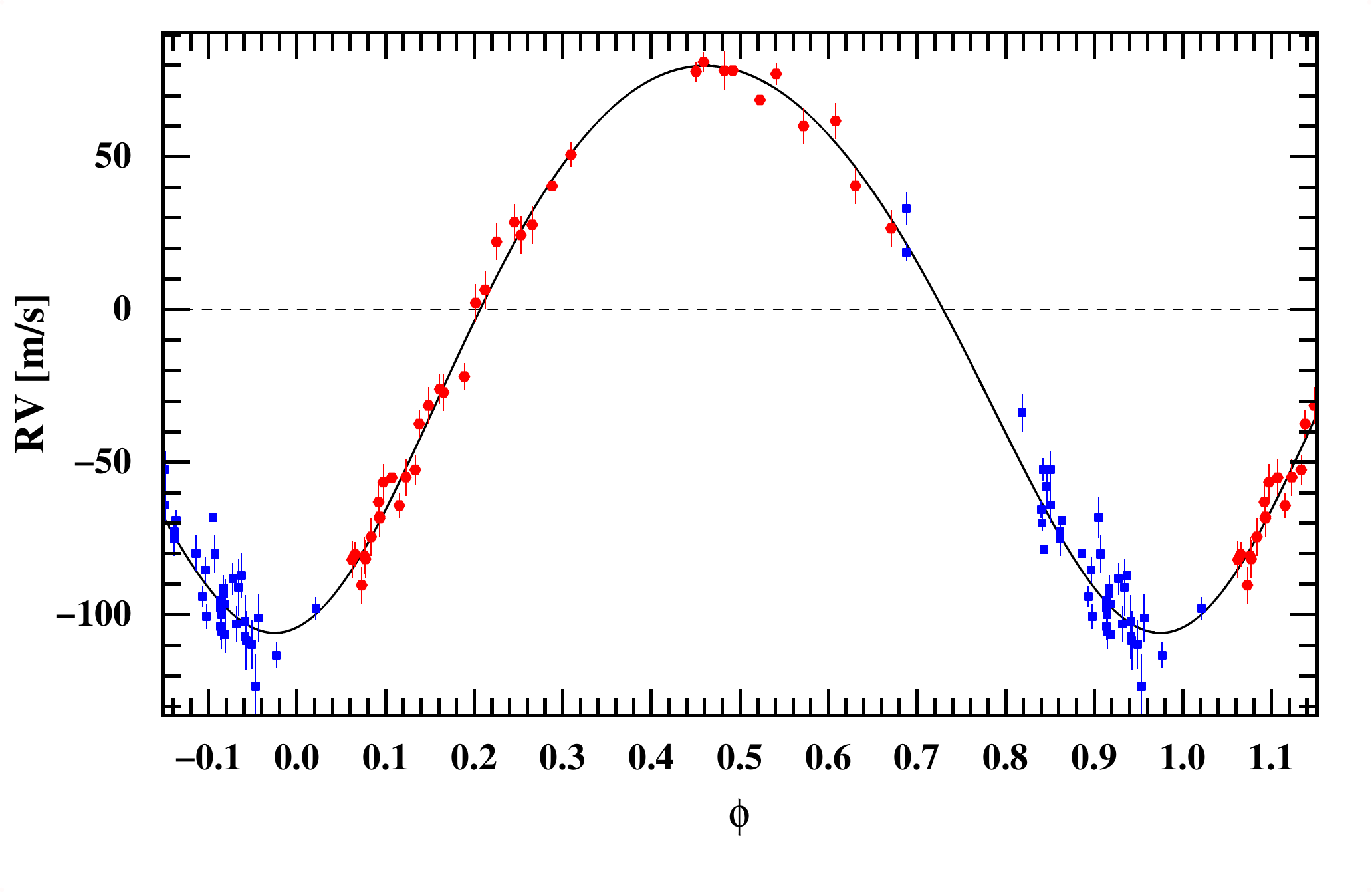,width=0.4\textwidth}
\caption{SOPHIE radial velocities and Keplerian model of the \a\, system: (top) as a function of time, with the residuals; (middle) as a function of phase for the inner planet; (bottom) as a function of phase for the outer planet. The blue points correspond to SOPHIE data obtained before June 2011,  the red points corresponds to more recent measurements after the SOPHIE+ upgrade.}
\label{rvafig}
\end{figure}

\begin{figure}[h!]
\begin{center}
\epsfig{file=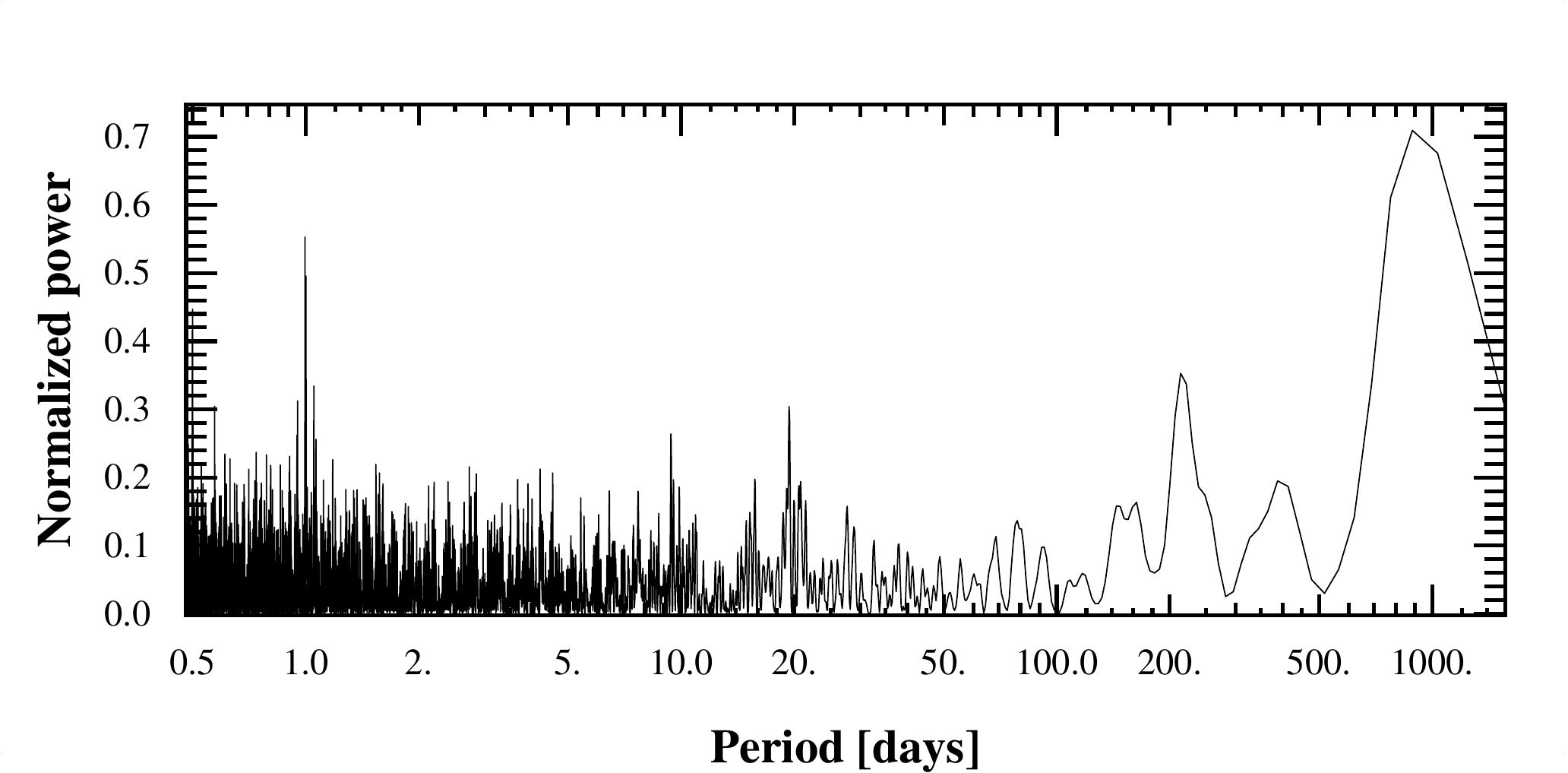,width=0.5\textwidth}
\epsfig{file=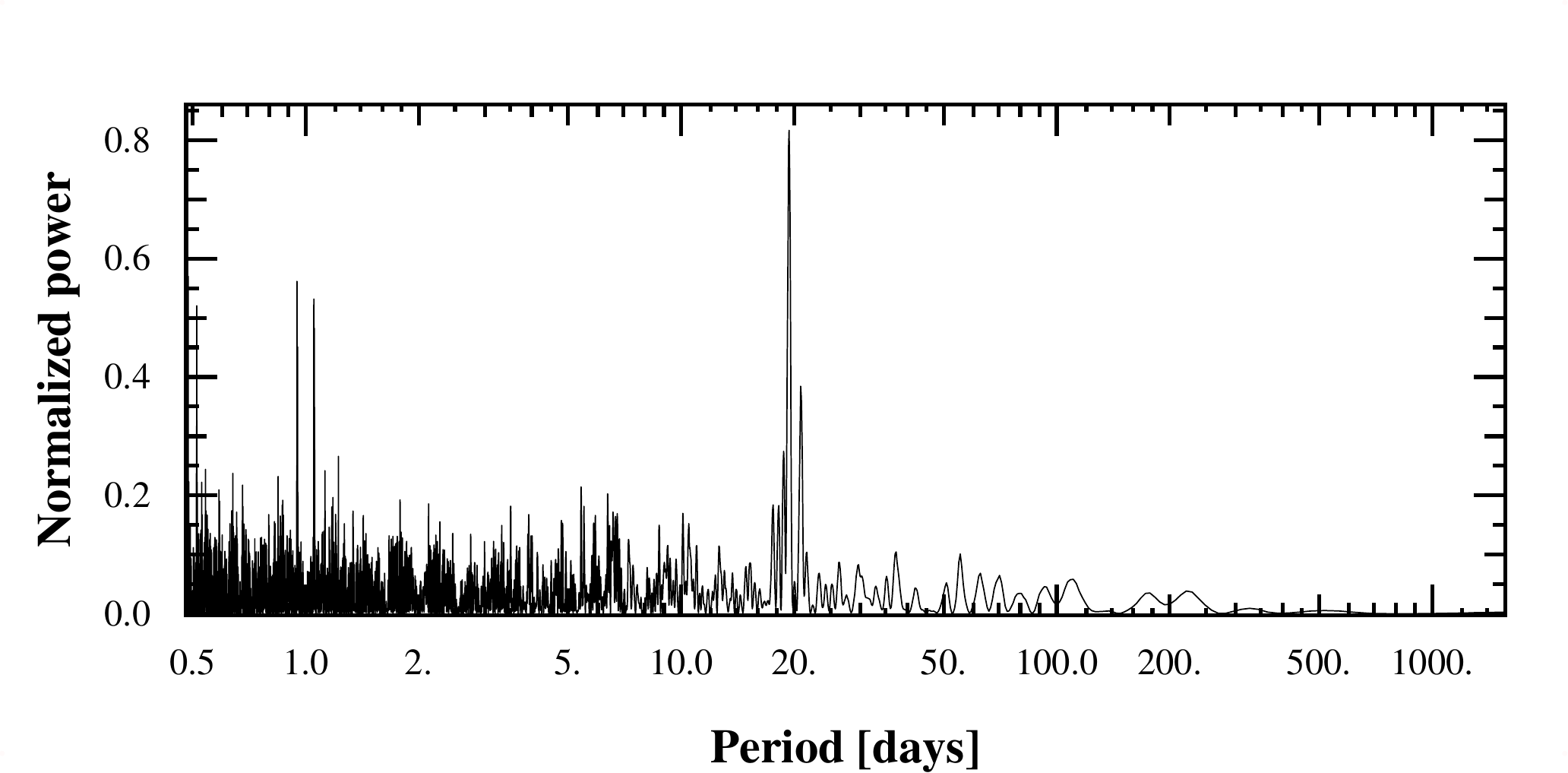,width=0.5\textwidth}
\epsfig{file=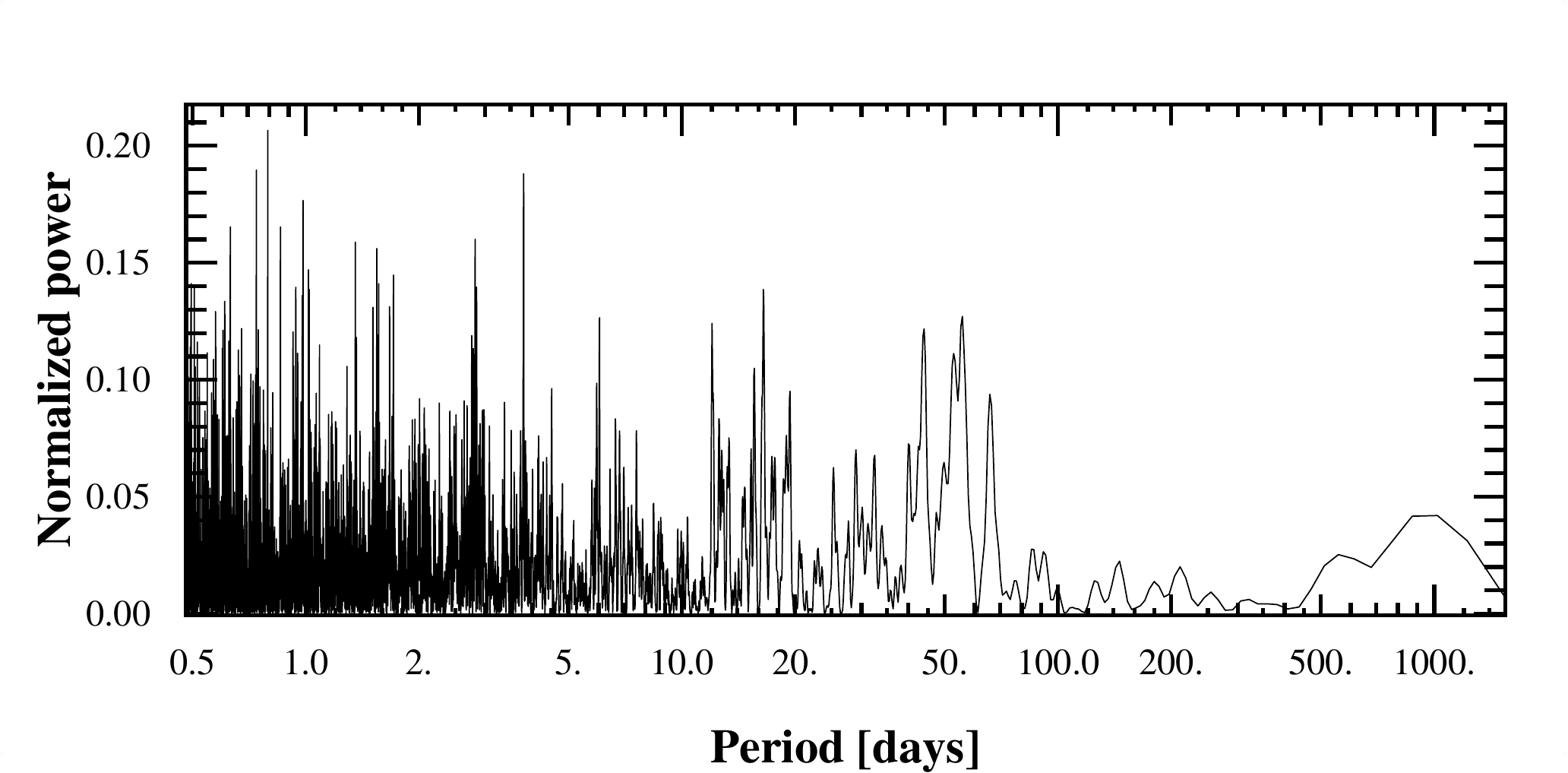,width=0.5\textwidth}
\caption{Periodogram of the radial velocity measurements of \a\ with both signals (top), with the outer-planet signal removed (middle), and periodogram of the bisector span (bottom). }
\label{perioa}
\end{center}
\end{figure}

\begin{table}
\tiny
\caption{Observed and inferred stellar parameters for the planet-hosting stars.  Uncertainties are noted between brackets. The horizontal line separates the parameters given in \citet{vanleeuwen07} and the parameters derived in this paper.}
\label{TableStars}
\centering
\begin{tabular}{l l c c c}
\hline\hline
\multicolumn{2}{l}{\bf Parameter} & \bf \a & \bf \b & \bf \c \\
\hline 
Sp & & F8V & G0V & G5V \\
$V$ & [mag] & 7.51 (0.01) &8.65 (0.01)  &8.65 (0.01)  \\
$B-V$ & [mag] & 0.53  & 0.54& 0.80 \\
$\pi$ & [mas] & 14.05 (0.70) & 14.45 (1.14) & 22.26 (0.66)  \\
$d$ & [pc] & 71.2 (3.5) & 69.2 (5.5) & 44.9 (1.3)\\
\hline
$M_V$ & [mag] & 3.25& 4.45& 5.39 \\
$B.C.$ & [mag] & -0.016 & -0.030& -0.134\\ 
$L$ & [$L_{\odot}$] & 4.0 & 1.3 & 0.6 \\
$T_{\mathrm{eff}}$ & [K]   &  6255 (66) & 6123 (65)&5519 (70)    \\
log $g$            & [cgs] &4.11 (0.11) &4.55 (0.11) &4.53 (0.12)  \\
$\mathrm{[Fe/H]}$  & [dex] &0.01 (0.04) & 0.05 (0.04)  & 0.23 (0.05)\\
$M_*$ & [M$_{\odot}$]  & 1.29 (0.04)&  1.125 (0.03)& 0.95 (0.03) \\
$v\sin{i}$ & [km s$^{-1}$] & 4.2 (0.5) &3.8 (0.5)  & 3.5 (0.5) \\
\rhk 		&			&  -4.9 (0.2) &-4.65 (0.1) & -4.65 (0.1) \\
%$P_{\mathrm{rot}}$ & [days] & &  &  \\
Age &[Gy]                   &  2.9 (0.4) &  1.25 (1.1) &  2.4 (2.4) \\
$R_*$ & [R$_{\odot}$]  & 1.67 (0.1)& 1.12 (0.05) &  1.0 (0.04) \\
\hline
\end{tabular}
\end{table}

\section{Stellar properties}
\label{stars}
The SOPHIE spectra were used to spectroscopically determine stellar parameters. The method described in \citet{santos} and \citet{sousa} was used on the mean high-signal-to-noise-ratio spectrum obtained for each star. We derived the effective temperature \teff, gravity \logg, and the iron content of the stellar atmospheres of \a, \b, and \c\ using equivalent width measurements of the Fe I and Fe II weak lines by imposing excitation and ionization equilibrium assuming local thermal equilibrium. Errors were obtained by quadratically adding respectively 60 K, 0.1, and 0.04 dex to the internal errors on \teff, \logg, and [Fe/H].  

The new reduction of the Hipparcos data \citep{vanleeuwen07} was used to obtain the stellar parallaxes and derive the stellar luminosities after including the bolometric corrections proposed by \citet{flower96} for the given effective temperatures. Then, the stellar mass, radius, and ages were derived using the evolutionary tracks and Bayesian estimation of \citet{dasilva06}. To take into account systematic effects, we adopted conservative uncertainties of 10\% on the stellar mass. 

The results are summarized in Table \ref{TableStars}. All three stars are solar-like dwarfs. \a\ and \b\ have solar metallicity, while \c\ has a significant excess of metal compared with the Sun, with [Fe/H]=0.23$\pm$0.05.

The cross-correlation function  allows various stellar indicators to be estimated: the bisector span (tracer of activity and blended stellar systems) and the projected rotational velocity (using calibrations developed in \citet{boisse10}). In addition, the activity level of the stellar chromosphere was estimated from the calcium H and K equivalent widths. The average value of the  \rhk\  parameter is also given in Table \ref{TableStars}. While \a\ is not active (\rhk $=$ -4.9 dex), \b\ and \c\ have a relatively high activity level (\rhk $=$ -4.65 dex for both). The variations of the \rhk\ value with time were systematically checked (Figure \ref{logr}) to verify whether the slow radial-velocity modulation is due to the long-term cycles of the star \citep{dumusque}. Then, the bisector variations were inspected to determine whether they correlate with the radial-velocity signal at each detected period. A negative correlation between the bisector spans and the radial velocities usually indicates that the distortion of stellar lines due to spot activity produces radial-velocity jitter \citep{queloz01, boisse12}, and the planet interpretation may be compromised. The expected jitter due to activity is of the order of 9, 12, and 12 \ms\,Êfor \a, \b, and \c, respectively, using calibrations described in \citet{santos00}. Finally, ranges for the rotation periods can be estimated from the $B-V$ and \rhk\ values. We found ranges of 5-9, 3-16, and 17-30 days, using the equations in \citet{noyes84} and a 3-$\sigma$ error range on \rhk.

\begin{figure}
\begin{center}
\epsfig{file=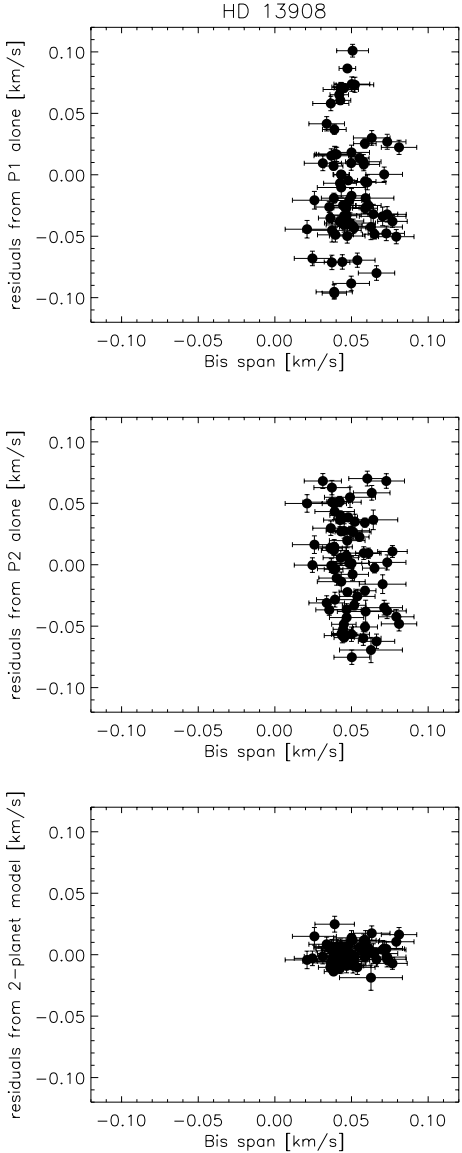,width=0.7\textwidth}
\caption{Bisector variations as a function of radial-velocity residuals from \a\ SOPHIE measurements, (top) without the outer planet signal, (middle) without the inner planet signal, and (bottom) without both planet signals.}
\label{bisa}
\end{center}
\end{figure}

\begin{figure}[h!]
\epsfig{file=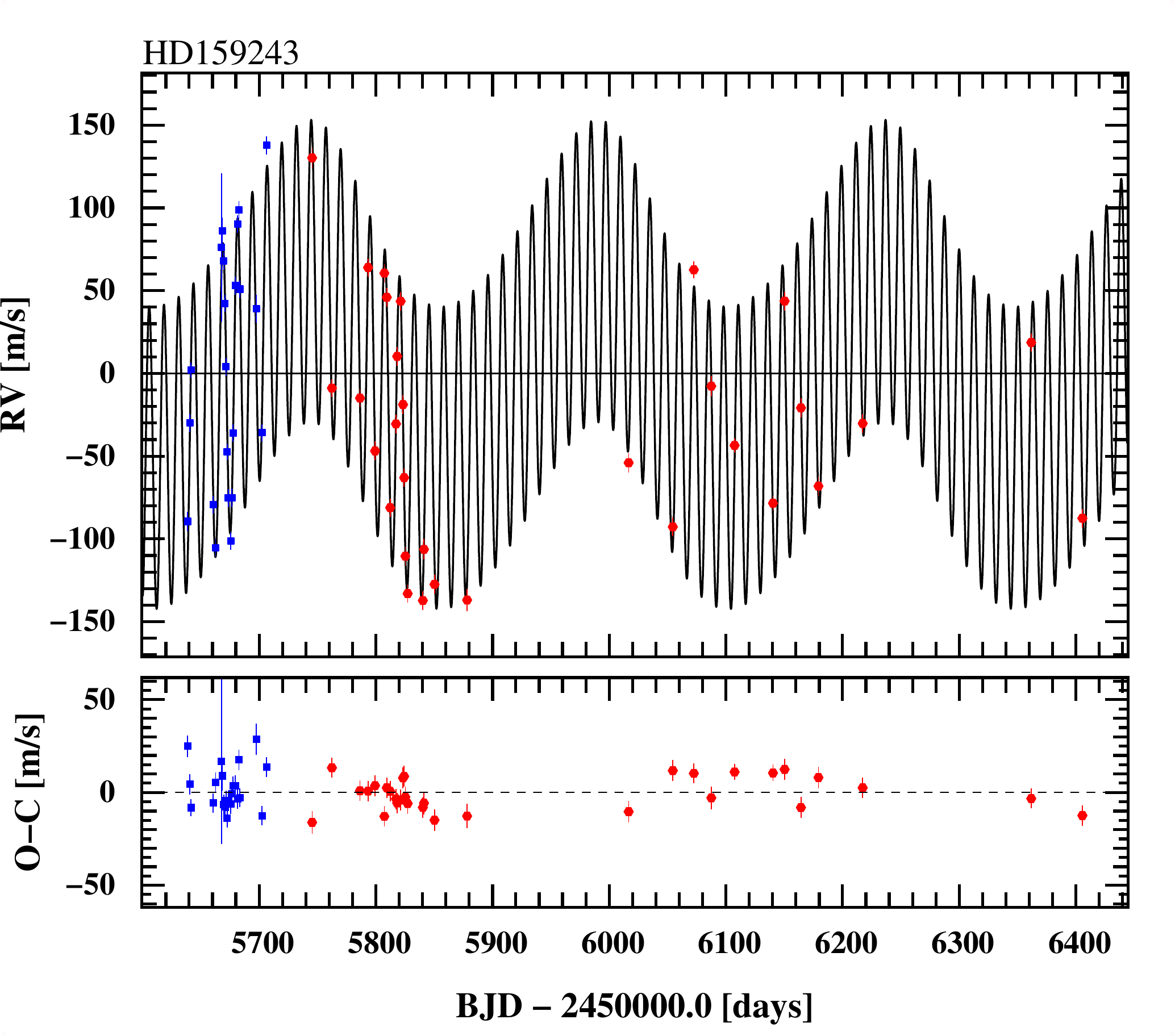,width=0.4\textwidth}
\epsfig{file=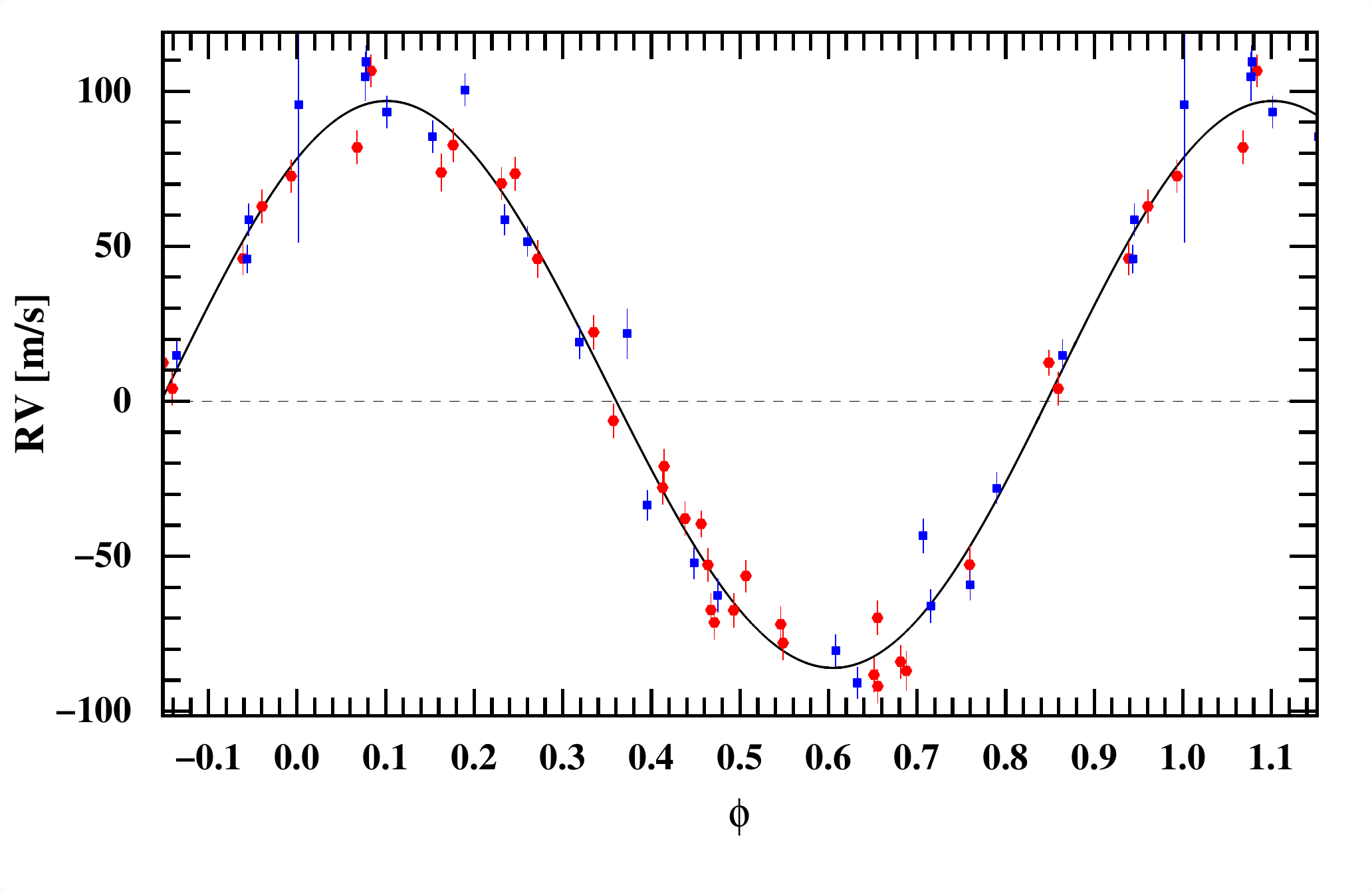,width=0.4\textwidth}
\epsfig{file=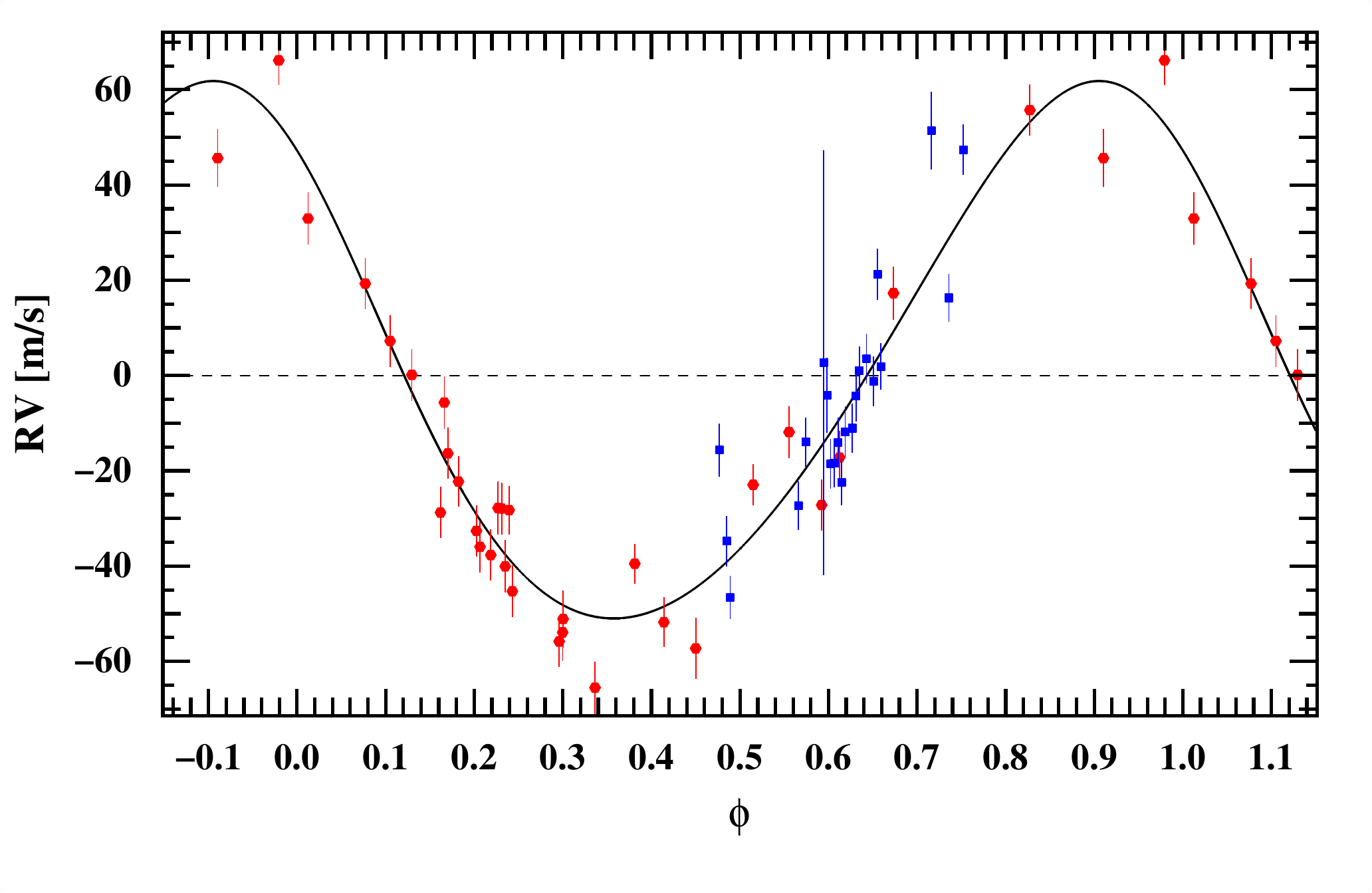,width=0.4\textwidth}
\caption{SOPHIE radial velocities and Keplerian model of the \b\, system. The display is the same as in Fig. 2, i.e. the middle and bottom panel correspond to the inner and outer planet, respectively.}
\label{rvbfig}
\end{figure}

\begin{figure}
\begin{center}
\epsfig{file=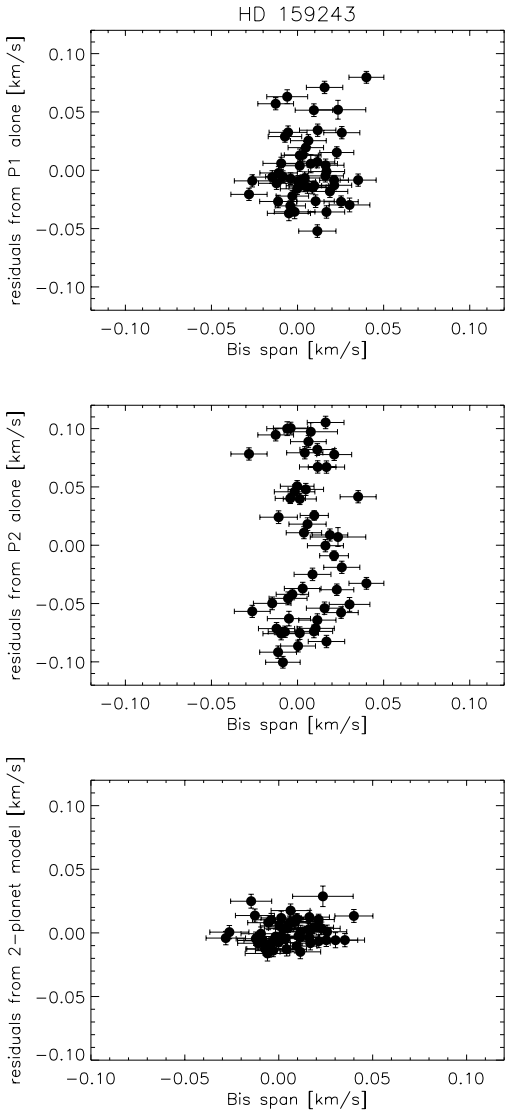,width=0.7\textwidth}
\caption{Bisector variations as a function of radial-velocity residuals for  \b. Same plots as in Fig. \ref{bisa}.}
\label{bisb}
\end{center}
\end{figure}

\begin{figure}[h!]
\begin{center}
\epsfig{file=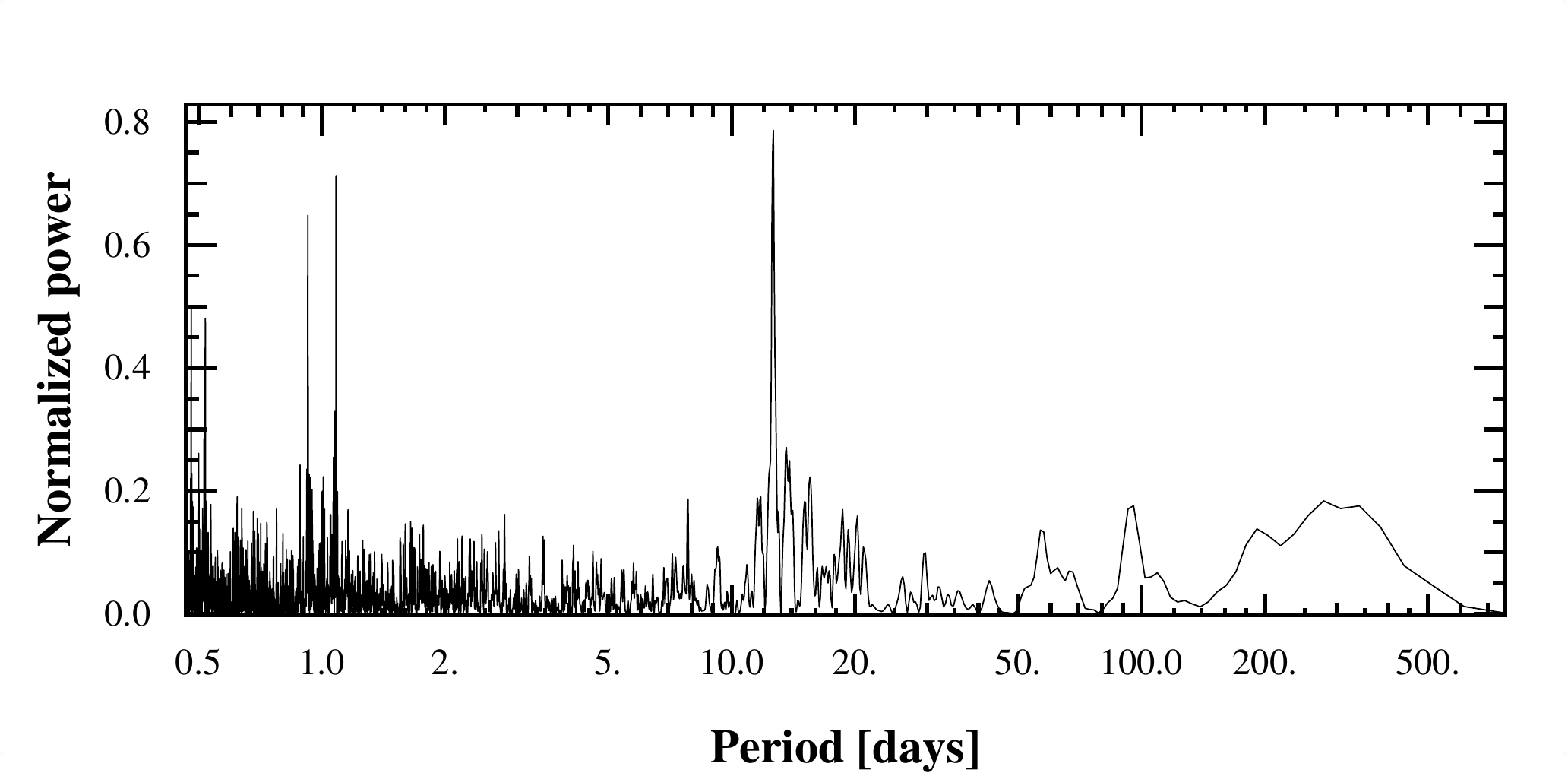,width=0.5\textwidth}
\epsfig{file=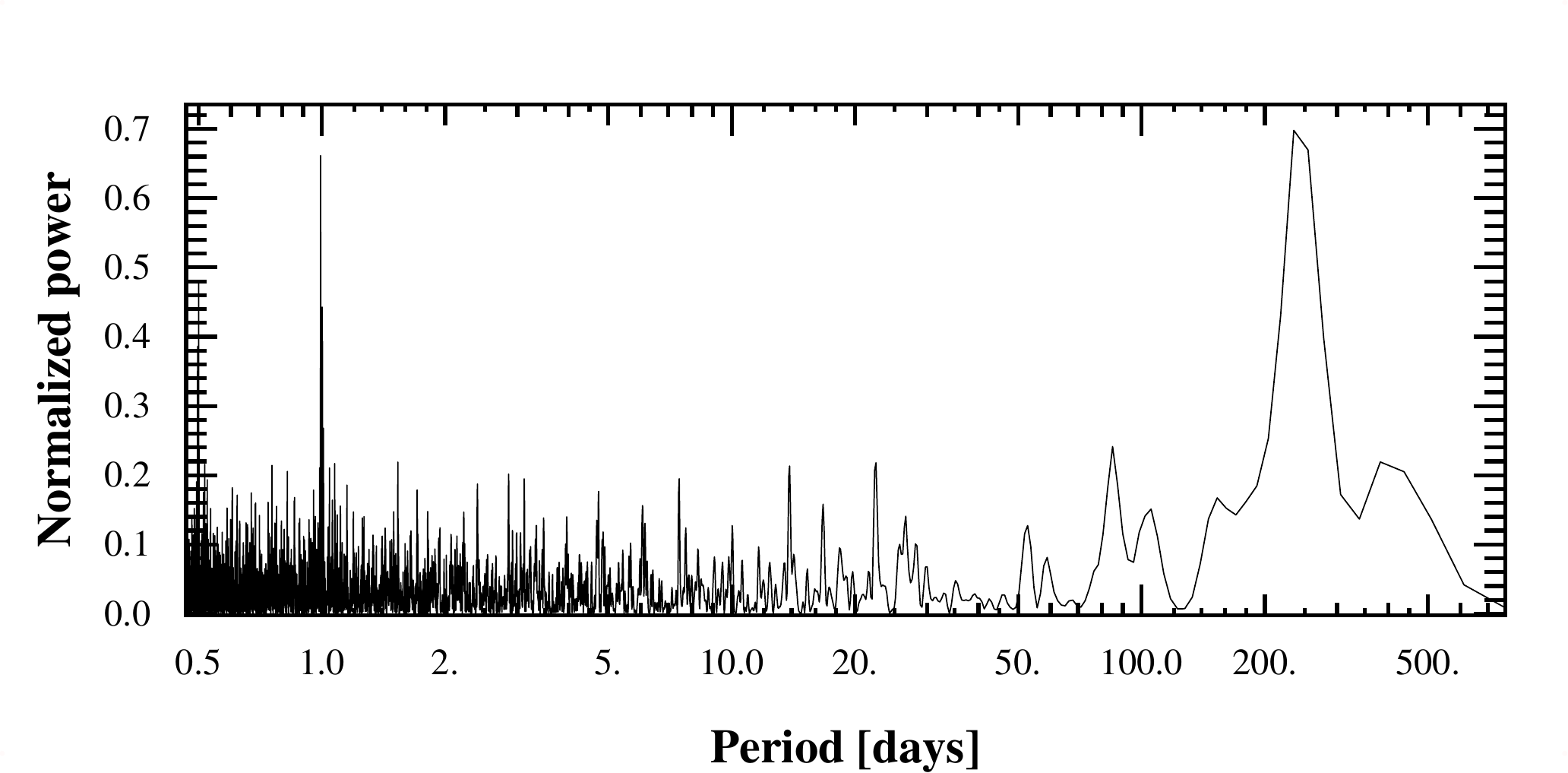,width=0.5\textwidth}
\epsfig{file=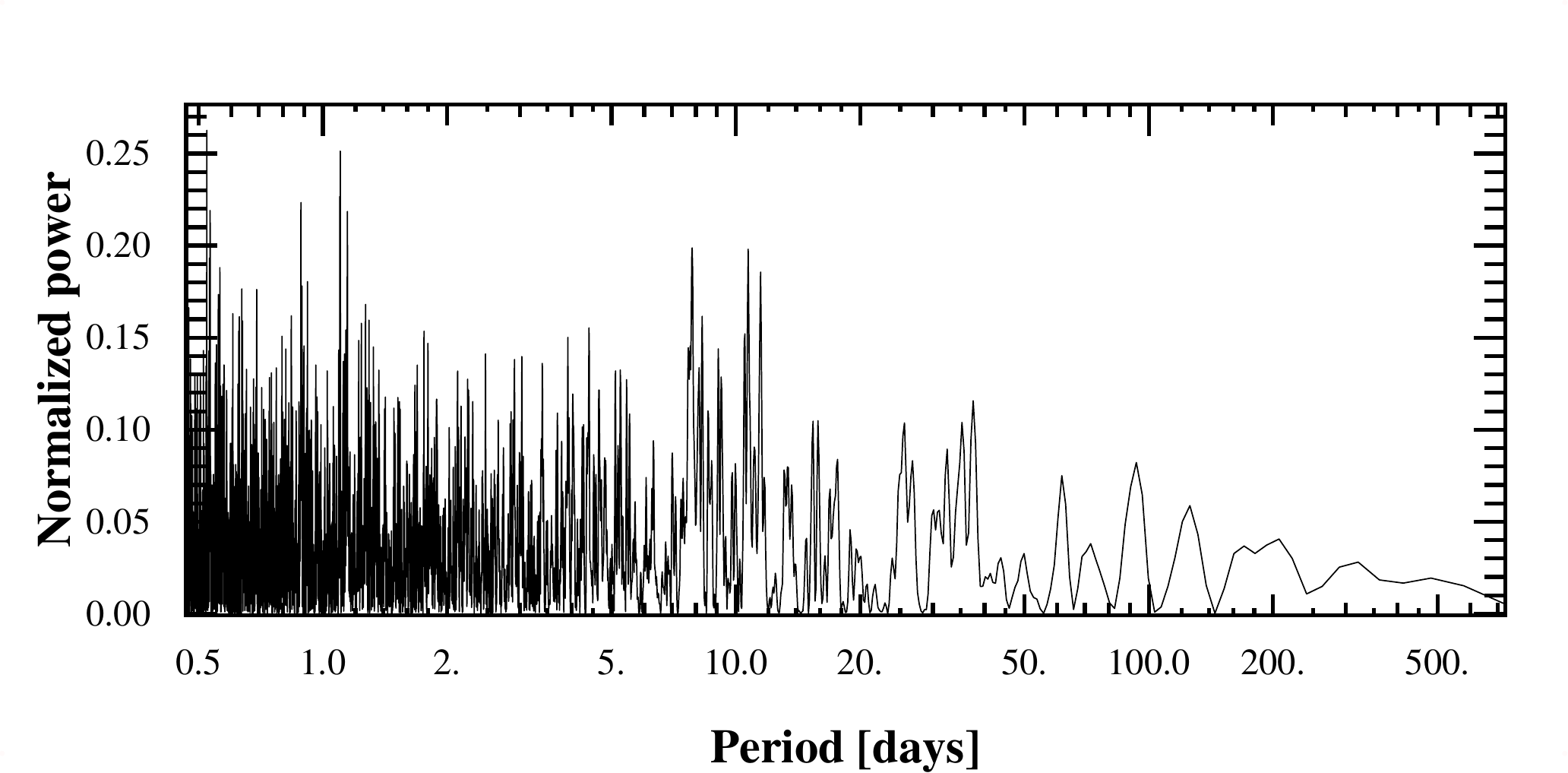,width=0.5\textwidth}
\caption{Periodogram of the radial-velocity measurements of \b\ with both signals (top), without the inner-planet signal  (middle), and the periodogram of the bisector span (bottom). }
\label{periob}
\end{center}
\end{figure}

\begin{figure}
\begin{center}
\epsfig{file=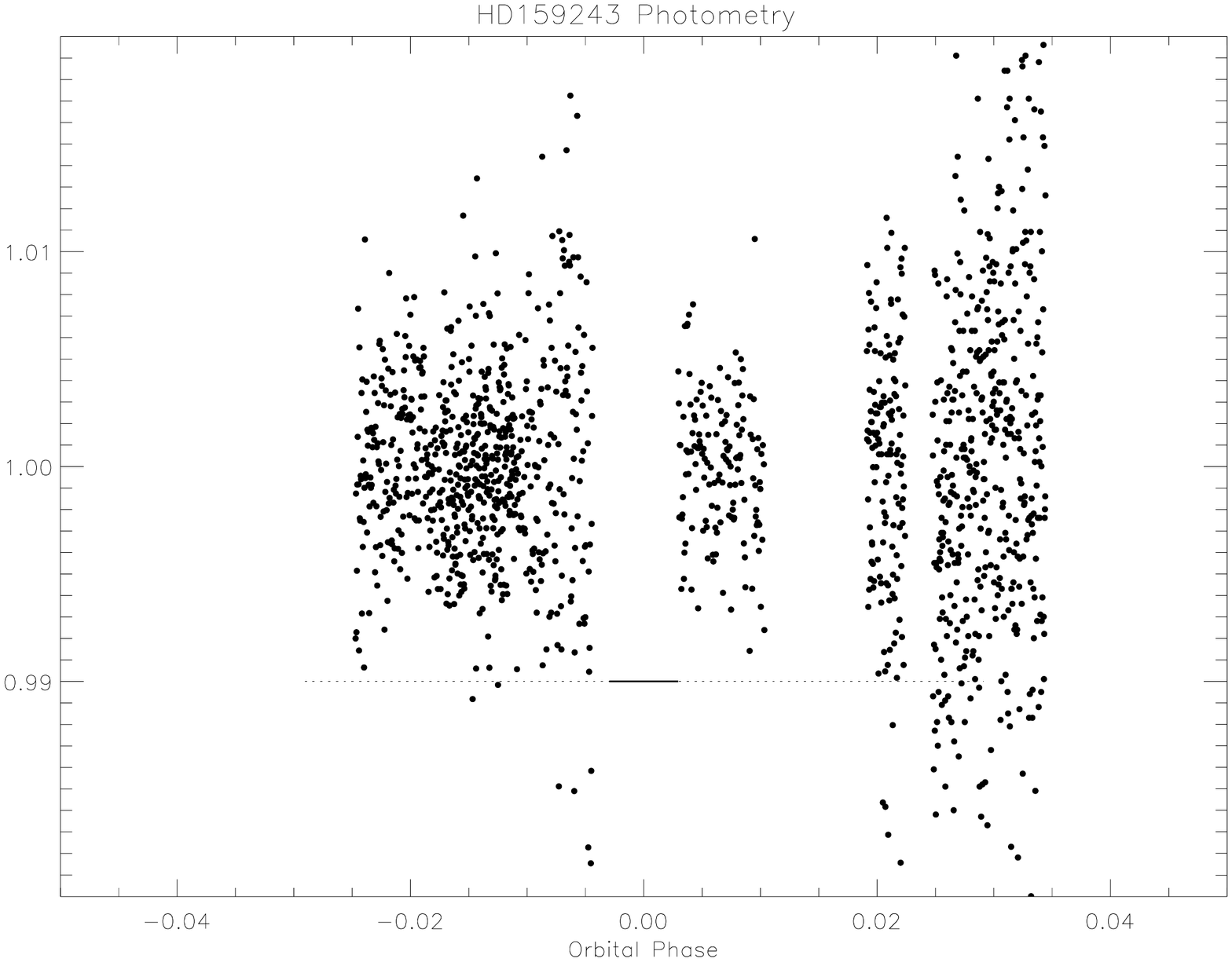,width=0.55\textwidth}
%x\vspace{-1cm}
\caption{ Relative flux of \b\ phased with the orbital period and time of transit of companion b. The dotted line shows the 1-$\sigma$ uncertainty in the transit ephemeris. The thick line shows the transit duration and expected transit depth for the nominal ephemeris. There is no evidence of a transit, although it cannot be ruled out in the whole range.}
\label{photomb}
\end{center}
\end{figure}

\section{Radial-velocity analysis}
\label{rv}
\subsection{\a}
Seventy-seven SOPHIE measurements have been collected on the star \a\ during more than four years between October 2008 and January 2013. The average noise of individual measurements is 7.9 \ms\ using SOPHIE, and 5.2 \ms\ using SOPHIE$+$ (Fig. \ref{rvafig}). The radial-velocity fluctuations have a standard deviation of 74 \ms, with a short-term component and a longer-term component. Because the measurements were secured in both the SOPHIE and SOPHIE+ configurations (Table \ref{rva}), we split the data set into two parts (delimited by the horizontal line in the table). 

Several solutions were tried and compared by estimating the standard deviation of the residuals:  single-planet model, single-planet model with linear and quadratic trend, two-planet model, two-planet model with trends, and three-planet models. The best solution was found with a two-Keplerian model with respective semi-amplitudes of 55.3$\pm$1.2 and 90.9$\pm$3.0 \ms (Fig. \ref{rvafig}).  The periodograms of the RV measurements are shown in Fig. \ref{perioa}. The peak corresponding to the outer companion is shown in the top plot; when the long-period signal is removed from the data, the inner-planet peak becomes proeminent (middle plot). The residuals of the two-planet model have an r.m.s. of 9.6 \ms\ for the first data set, and 5.3 \ms\ for the second data set after the spectrograph's upgrade. These residual values are close to the one predicted from the activity jitter of 9 \ms\ in \citet{santos00} and may be attributed to the stellar photospheric activity.

We searched in vain for periodicity in the time series of the bisector span (Fig. \ref{perioa} bottom) and the chromospheric index $\log R'_{\mathrm{HK}}$, therefore it seems unlikely that the short-period signal is generated by the rotation period of the star or that the long-period signal is related to the activity cycle of \a\ \citep{dumusque}. The bisector spans do not vary with the radial velocity for any of the detected signals, which rules out that the photospheric activity is the main source of the variations (Figure \ref{bisa}). The origin of the two signals is most likely the Doppler shift of the stellar spectrum that is caused by the orbital motion of the planets around the star and not by the deformation of the stellar profile due to photospheric activity. Note that the scatter of the bisector and \rhk\ activity indices are 13 \ms\ and 0.15 dex.

Because activity is discarded as a possible source of the radial-velocity variations, the system around \a\ is very likely composed of a Jupiter-like planet in a 19.4-day orbit and a massive giant planet in a 2.5-year long orbit. Parameters obtained with Monte Carlo Markov Chain simulations in {\sc yorbit}, using 500,000 iterations, are given in Table \ref{sola}. Although the eccentricity of the inner planet is not significantly different from zero, we kept it as a free parameter to include the associated error in the general modeling of the system.

The short-period planet has a transit probability of 3\%. Photometric observations were attempted. No transit was detected because we missed the transit window, as we realized later from the refined ephemeris. Additional observations are therefore encouraged at the transit times given by $JD = 2,455,756.032+N \times 19.382$.

Finally, we searched the Hipparcos astrometry data for the signature of the outer companion with the method presented in \citet{sahlmann11}.  Since five orbital elements are determined by spectroscopy (P, e, T$_0$, $\omega$, K), the intermediate-astrometry data \citep{vanleeuwen07} can constrain the two remaining orbital parameters, in particular the orbit inclination $i$ that yields the companion's mass. No astrometric signature corresponding to the spectroscopic orbit was detected, which can be translated into a limit on $i$ and an upper limit of 0.15 \Msun\ for the mass of \a\ c. This constraint is compatible with the non-detection of a secondary signal in the cross-correlation function.

\begin{table} 
\caption{Orbital and physical parameters for the planets orbiting the star \a. 
$T$ is the epoch of the highest RV. $\sigma$(O-C) is the residual noise after orbital fitting.  }
\label{TablePlanets}
\centering
\begin{tabular}{l l c c c }
\hline\hline
\multicolumn{2}{l}{\bf Parameter}&
& \bf \a\,b & \bf \a\,c \\
\hline
$P$ & [days] & 							&19.382 $\pm$0.006& 931$\pm$17 \\
$T$ & [JD-2400000] & 					&55750.93$\pm$0.18 &56165$\pm$9 \\
$T_{transit}$& [JD-2400000] & 				&55756.032$\pm$0.156&56436$\pm$15\\
  $e$ &            &  						&0.046$\pm$0.022& 0.12$\pm$0.02 \\
$\omega$ & [deg]    & 					&185$\pm$43&185$\pm$11\\
$K$ & [m s$^{-1}$] &   					&55.3$\pm$1.2& 90.9$\pm$3.0 \\
$m_2 \sin{i}$ & [M$_{\mathrm{Jup}}$] & 		&0.865$\pm$0.035& 5.13$\pm$0.25 \\
$a$ & [AU] &   							&0.154$\pm$0.0025&2.03$\pm$0.04 \\
\hline					
$\gamma_{SOPHIE}$ & [km s$^{-1}$] &   			&15.893 $\pm$0.008& \\ 
$\gamma_{SOPHIE+}$ & [km s$^{-1}$] &   			&15.882$\pm$0.006& \\ 
$N_{\mathrm{meas}}$ & &				&77&\\
Span & [days]       &  						&1588.7& \\
$\sigma$ (O-C) & [m s$^{-1}$] & 			&9.6/5.3& \\
$\chi^2_{red}$& & 						&2.6& \\
\hline
\label{sola}
\end{tabular}
\end{table}

\begin{figure}[h!]
\epsfig{file=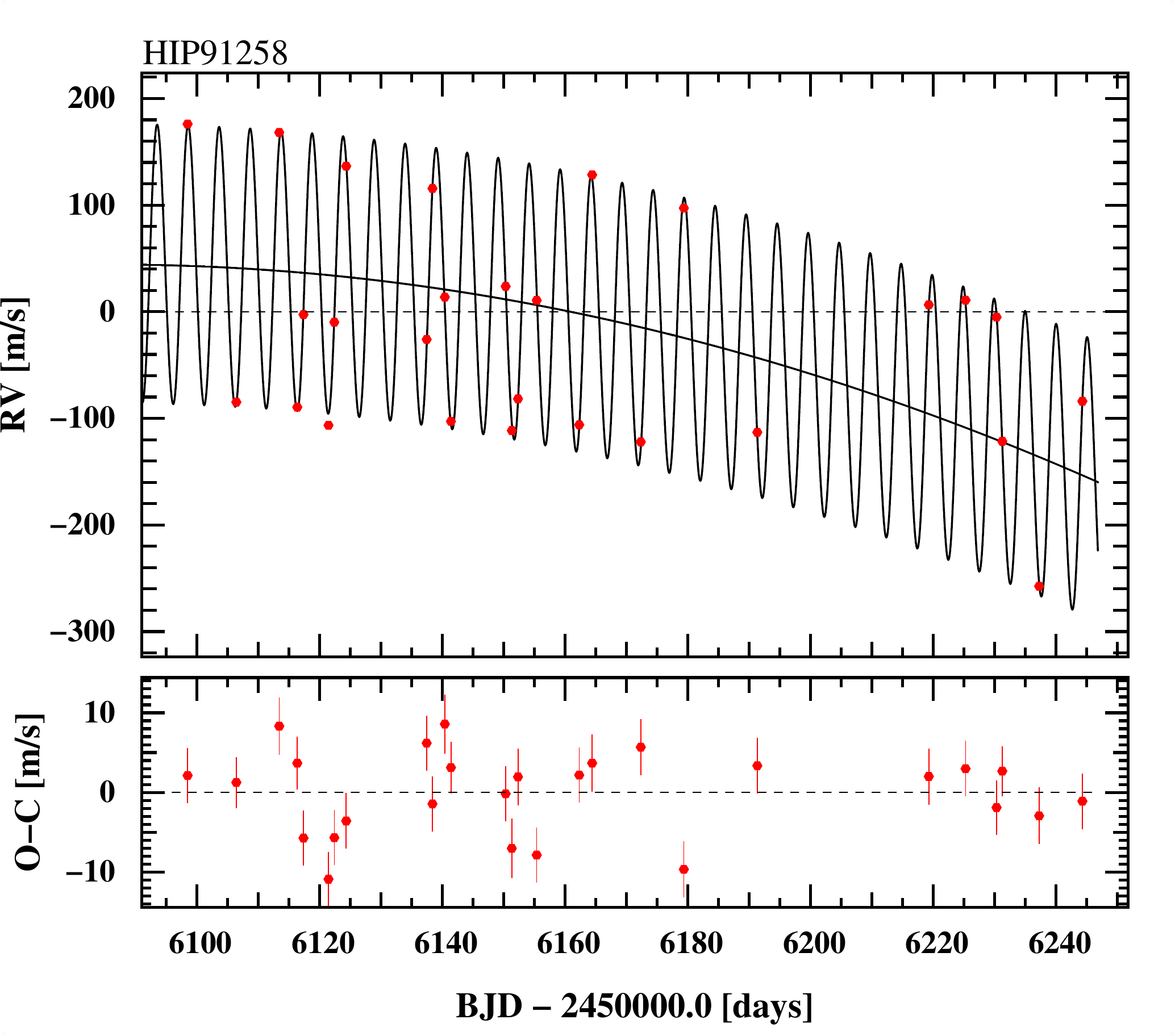,width=0.4\textwidth}
\epsfig{file=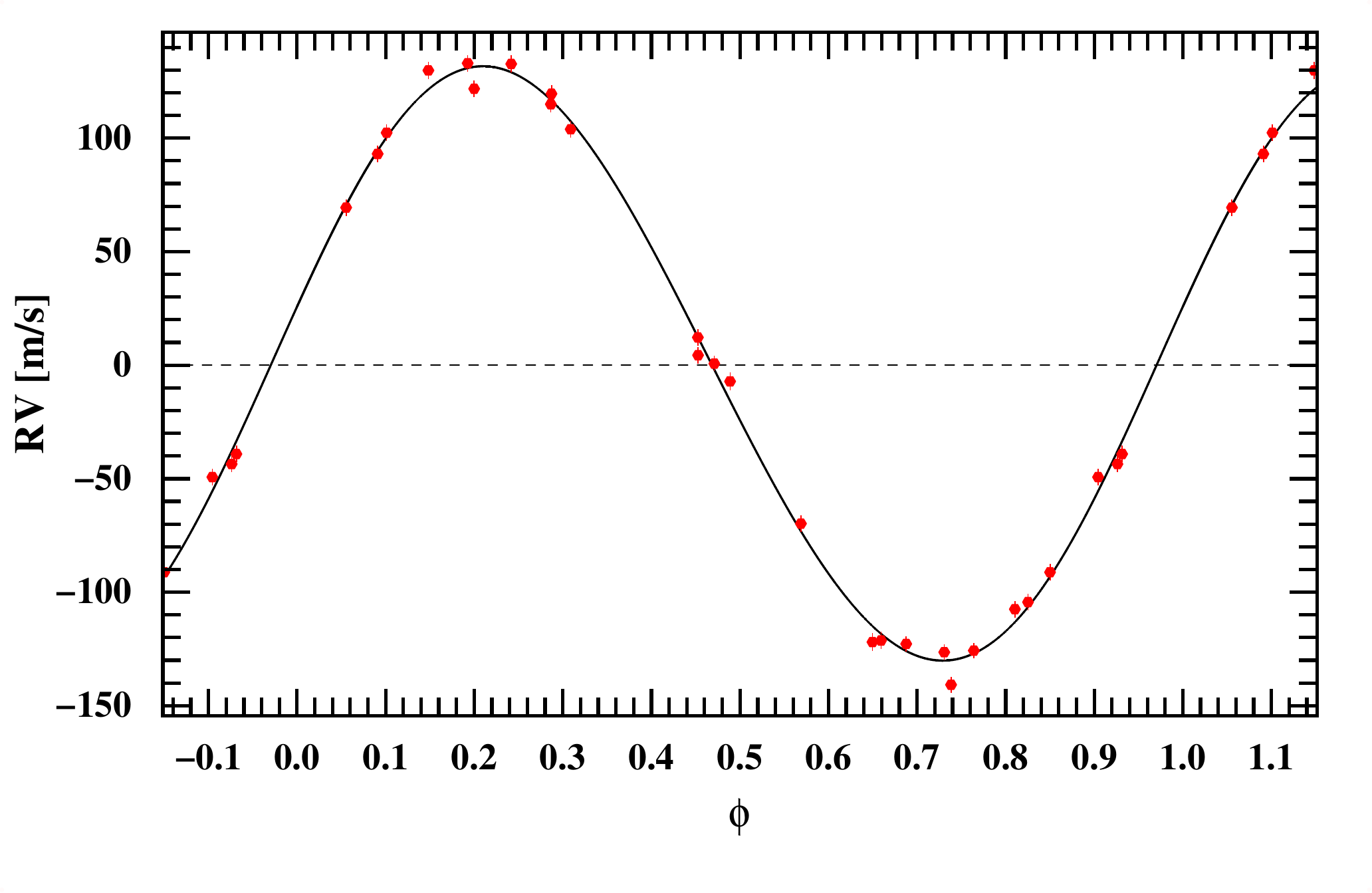,width=0.4\textwidth}
\caption{SOPHIE radial velocities and Keplerian model of the system orbiting \c: (top) as a function of time, with the residuals to the model including a quadratic long-term trend, and (bottom) as a function of phase for the inner planet.   }
\label{rvcfig}
\end{figure}

\begin{figure}[h!]
\begin{center}
\epsfig{file=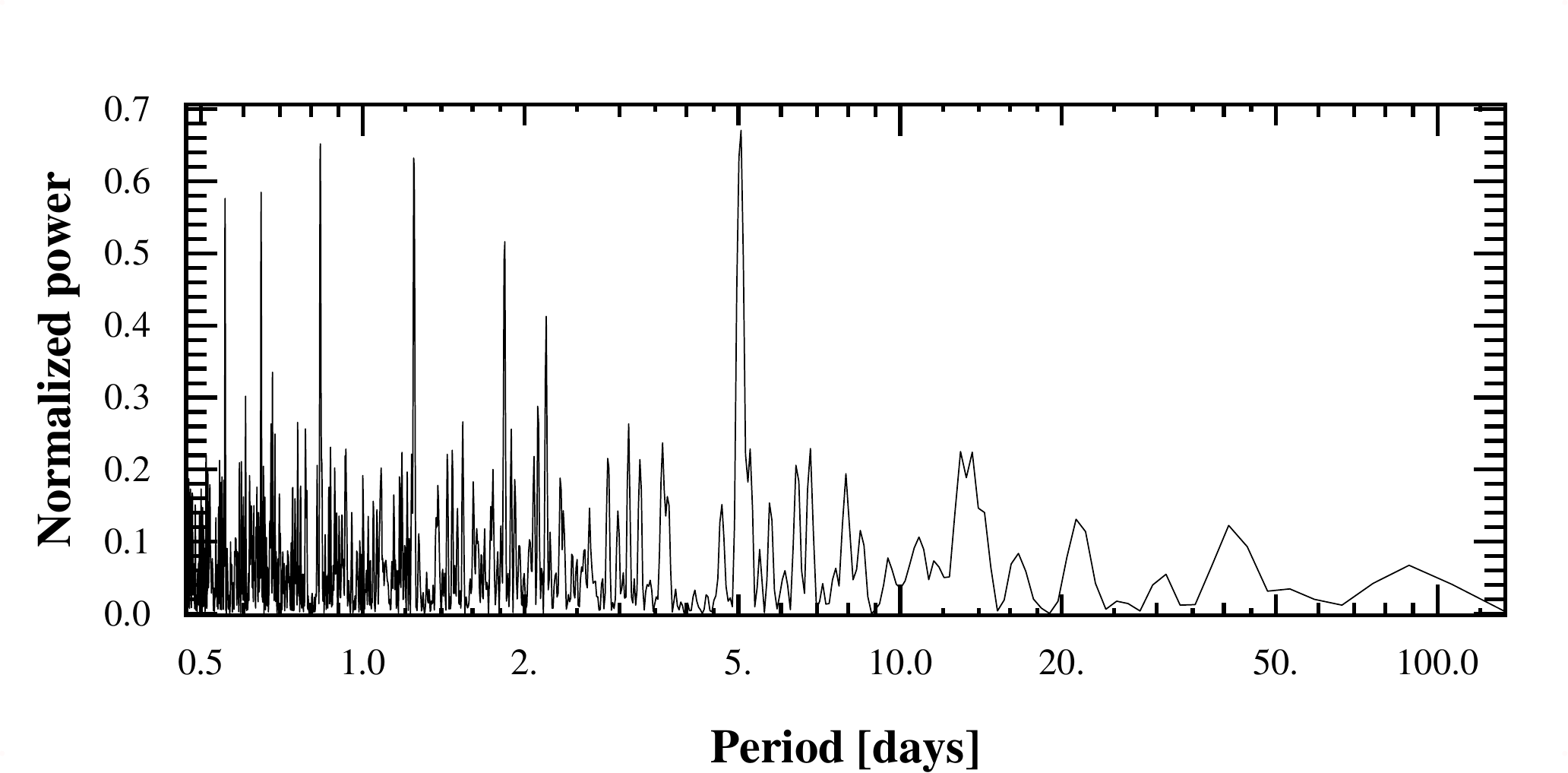,width=0.5\textwidth}
\epsfig{file=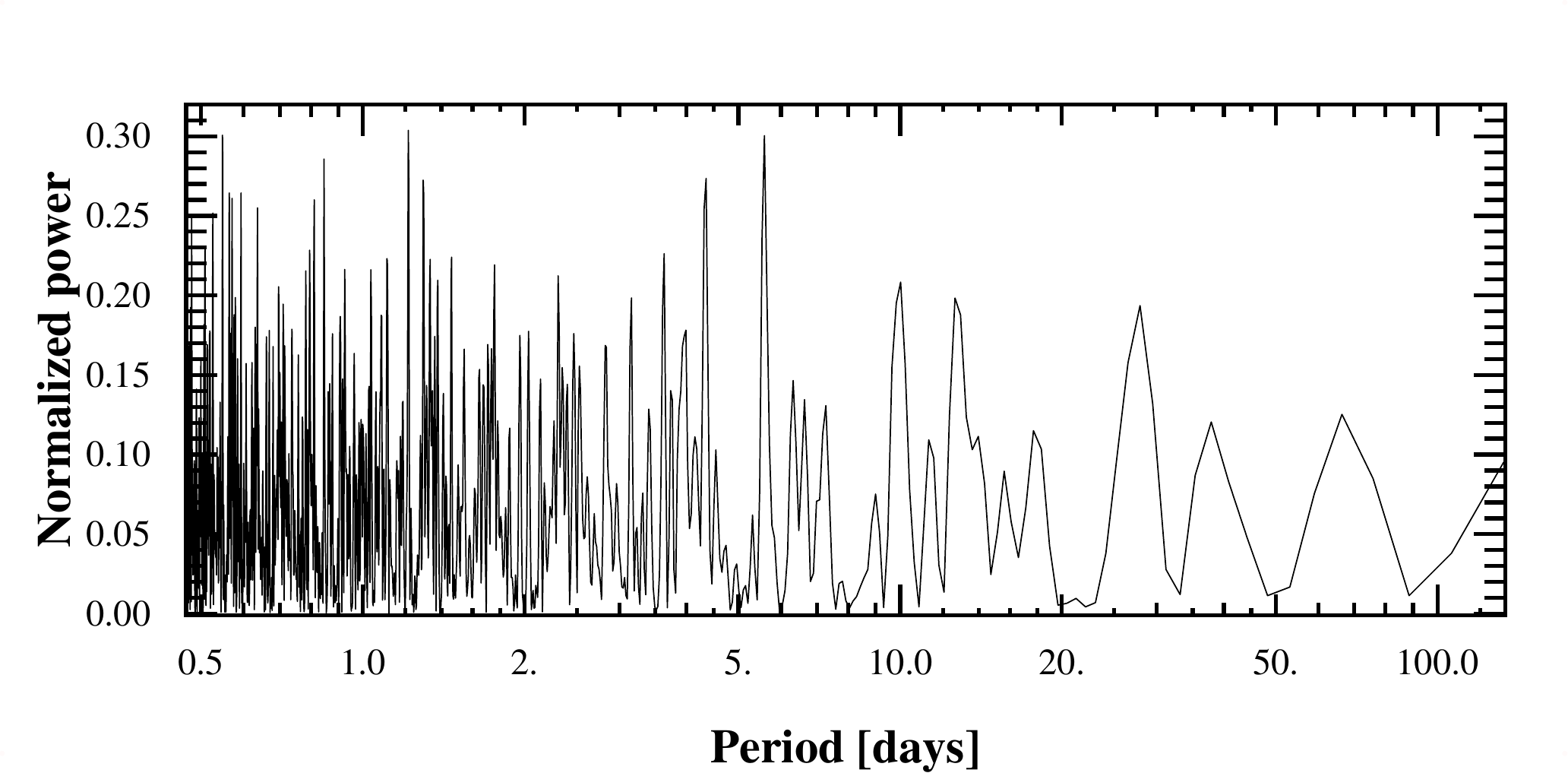,width=0.5\textwidth}
\caption{Periodogram of the radial-velocity measurements of \c\  (top) and of the bisector span (bottom). }
\label{perioc}
\end{center}
\end{figure}

\begin{figure}
\begin{center}
\vspace{-1.5cm}
\epsfig{file=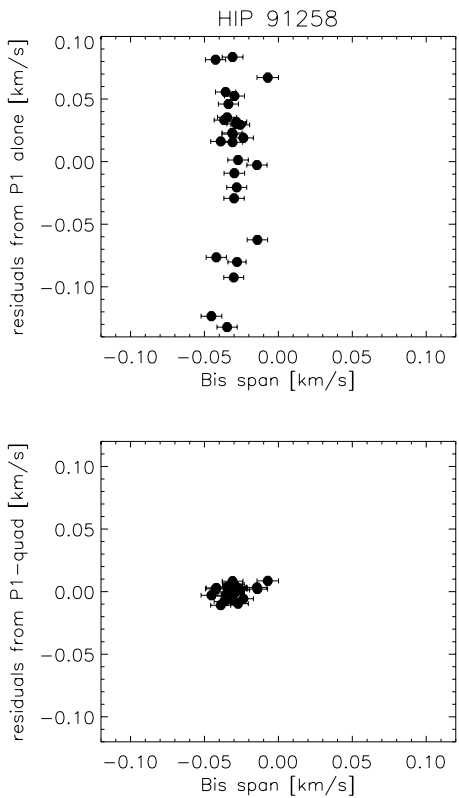,width=0.7\textwidth}
\vspace{-3.5cm}
\caption{Bisector variations as a function of radial-velocity residuals for \c. (top) without the 5.05-day period planet signal, and (bottom) without both the planet signal and the quadratic trend. }
\label{bisc}
\end{center}
\end{figure}

\subsection{\b}
We secured fifty-three SOPHIE radial-velocity measurements of  \b\, during 767 days, from March 2011 to April 2013. Thirty-two measurements were obtained after the SOPHIE+ upgrade. The total data set features individual error bars with an average of 5.2 \ms\ (Fig. \ref{rvbfig}). The standard deviation of the time series is 71 \ms. In the same way as for \a, we compared several models with sets of 1, 2, or 3 planets and trends. The best-fit model was obtained with two Keplerian orbits, although it gives a residual noise of 12.4 and 9.4 \ms\ for the two data sets. 

The amplitude of this jitter is compatible with the expected jitter of 12 \ms\ (see Section \ref{stars}). It is therefore probably caused by activity, since the average \rhk\ is -4.65 dex, indicative of an active star. During the period of observations, however, the stars showed almost no variation of its bisector span (r.m.s.=15\ms) and chromospheric activity (Fig. \ref{logr}). The \rhk\ parameter has an r.m.s. of 0.08 dex, and its variation is not correlated with the radial velocity signal. The radial-velocity variations are only weakly correlated with the bisector span for the two signals at 12.6 and 248 days (Fig. \ref{bisb}). The estimated rotation period of $\simeq$10 days (see Section \ref{stars}) and the marginally correlated behaviour of the bisector span without the outer planet signal (Fig. \ref{bisb} middle), however, might cast some doubt on the inner planet. With an amplitude of 91 \ms, it is nevertheless unlikely that the signal is entirely generated by spot activity.  With an activity-induced scatter amplitude of the order of 12 \ms, spot-related activity could  explain the residual noise level, but not the 12-day period or the 91-\ms\ amplitude signal. Dedicated activity simulations are needed to estimate the required correction, as in \citet{boisse12}. In Figure \ref{periob}, we compare the periodogram of the radial-velocity measurements with the periodogram of the bisector spans. While the former (top plot) exhibits a strong peak at 12.6 days, the latter (bottom plot) shows faint peaks at 7.8 and 10.9 days. In this paper, we neglected the contribution of the activity to the inner-planet signal. 

The best two-Keplerian model shows that the system of \b\, is composed of a 12.6-day Jupiter-like planet (minimum mass of 1.13$\pm$0.05 \Mjup) with a second giant planet with a minimum mass of 1.9$\pm$0.13 \Mjup\ in a 248-day period. The outer planet, despite its lower amplitude, is clearly detected in the periodogram of the radial-velocity measurements without the inner-planet signal (Fig. \ref{periob}, middle plot). The two planetary companions have circular orbits. Table \ref{solb} gives the parameters of the planets obtained after 500,000 MCMC iterations. The non-detection of an astrometric signal with Hipparcos sets an upper limit of 0.52 \Msun\ for the outer companion \b\,c , a weak constraint that can also be established independently by the non-detection of a secondary signal in the spectra.
 
A transit search was carried out with several telescopes despite the relatively low transit probability of 4\%. Observations were carried out on 30 July 2012 at Oversky Observatory\footnote{http://www.over-sky.fr/} (La Palma, Spain) with a 14-inch telescope. The instrument includes a CCD SBIG STL-1001e camera and a Sloan $r'$ filter. Photometric observations were reduced using the software Muniwin 2.0. The sequence lasted 4.3 hours and the normalized flux reached a scatter  of 0.7\%. A second series of observations was secured on 9 April 2013 at Saint Michel l'Observatoire. It was obtained with a 20-cm telescope at f/5.5  equipped with an SBIG ST8XME camera during 2.2 hours; a standard deviation of the normalized flux 0.46\% was measured. Two additional sequences on 4 May and 10 June 2013  made use of the ROTAT\footnote{http://stargate-ohp.de/} 60-cm telescope at Haute-Provence Observatory, equipped with an SBIG STL11000 camera at the Newton focus reduced at f/3.2.  Images of all sequences were processed with Muniwin 2.0 and relative photometry was extracted using one or several comparison stars. The first sequence of 4.1 hour duration has a scatter of 0.5\% in relative flux; the second sequence lasted 3.9 hours and the flux has an r.m.s. of 0.6\%. Together the photometric observations cover 75\% of the transit window as allowed at the latest date of observations (June 2013). The transit ephemeris error ($\sim$9 hours) is now much larger than the transit duration (expected 1.8 hour). Recovering a more accurate transit ephemeris would require a new intensive radial-velocity campaign over several weeks. The standard deviation of the full photometric time series is 0.7\%, and no transit is detected with a limiting depth of about 0.5\%.  Figure \ref{photomb} shows the relative flux of \b\, and the expected ephemeris, duration, and depth of a potential transit, with a tolerance of 9 hours corresponding to the propagated error in mid-2013. More data are needed to conclude.

\begin{table} 
\caption{Orbital and physical parameters for the planets orbiting \b.  }
\label{TablePlanets}
\centering
\begin{tabular}{l l c c c }
\hline\hline
\multicolumn{2}{l}{\bf Parameter}&
& \bf \b\,b & \bf \b\,c \\
\hline
$P$ & [days] & 							&12.620$\pm$0.004&248.4 $\pm$4.9\\
$T$ & [JD-2400000] & 					&56426.11$\pm$0.21 &56484$\pm$11\\
$T_{transit}$& [JD-2400000] & 				&56416.78$\pm$0.22 &56428$\pm$102\\
  $e$ &            &  						&0.02$\pm$0.018&0.075$\pm$0.05 \\
$\omega$ & [deg]    & 					&223$\pm$93& 69$\pm$97\\
$K$ & [m s$^{-1}$] &   					&91.1$\pm$2.1&56.6 $\pm$3.3 \\
$m_2 \sin{i}$ & [M$_{\mathrm{Jup}}$] & 		&1.13$\pm$0.05& 1.9$\pm$0.13 \\
$a$ & [AU] &   							&0.11$\pm$0.002&0.80$\pm$0.02 \\
\hline					
$\gamma_{SOPHIE}$ & [km s$^{-1}$] &   			&-22.653 $\pm$0.012& \\ 
$\gamma_{SOPHIE+}$ & [km s$^{-1}$] &   			&-22.69$\pm$0.01& \\ 
$N_{\mathrm{meas}}$ & &				&53&\\
Span & [days]       &  						&767& \\
$\sigma$ (O-C) & [m s$^{-1}$] & 			&12.4/9.4& \\
$\chi^2_{red}$& & 						&4.2& \\
\hline
\label{solb}
\end{tabular}
\end{table}

\subsection{\c}
The star \c\ was observed with SOPHIE+ only after its major upgrade in June 2011. Twenty-seven measurements were secured on this target. The average uncertainty of this data set is 3.4 \ms. The periodogram of the RV series shows a peak at 5 days and its harmonics and aliases, as seen in Fig. \ref{perioc}. When modelled with a single Keplerian, we derive a standard deviation of the residuals of 45 \ms, indicating that another signal is present.  The residuals drop to 13 \ms\ when a linear trend is added, then to 6 \ms\ when the trend is quadratic. Adding a second planet to the fit does not improve the residuals. The simplest and best fit is thus obtained with a quadratic trend  and a short-period planet. The star does not show strong signs of activity, with a flat bisector span as a function of the radial velocity (Fig. \ref{bisc}) and an r.m.s. of 8 \ms. {The periodogram of the bisector spans shows only noise (Fig. \ref{perioc} bottom).} The \rhk\ value varies very little and shows no cyclic behaviour (Fig.Ê\ref{logr}), with a standard deviation of 0.03 dex during the period of observations. When the Keplerian+quadratric trend model is removed from the data, the periodogram shows a peak near 29 days that could be caused by activity, since the estimated stellar rotation period from the average \rhk\ index is 24$\pm$7 days. 

The main detected signal is attributed to a planet with a minimum mass of 1.09 \Mjup\, in a 5.05 day circular orbit with a semi-amplitude of 130.9$\pm$1.7 \ms. The full set of parameters is given in Table \ref{solc}.
When a model with two Keplerian orbits is adjusted to the data, the best solution favours an outer body with a  5-yr period and a mass slightly above the hydrogen-burning limit.  The time span of observations is, however, insufficient to precisely characterize of this distant body in the system of \c, and we defer its identification to future studies based on several-years long additional observations.  If adjusted to the shortest possible period of 150 days and using a null eccentricity and an RV amplitude of 100\ms\ as observed, the lowest possible mass of the outer companion is 2.5 \Mjup. A regular RV monitoring with five to ten measurements per year over two years would provide a sufficient constraint to determine whether the outer companion is in the planet, brown dwarf or star domain. Adaptive-optics imaging or future space astrometry with GAIA of this target would also allow  constraining a possible stellar companion.

\begin{figure}
\begin{center}
\epsfig{file=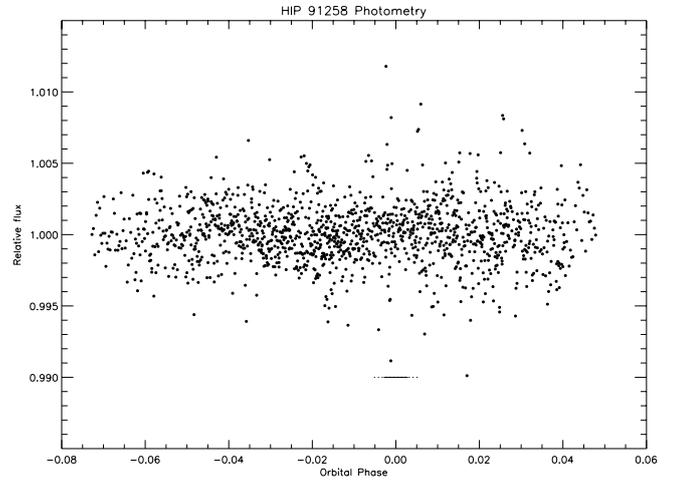,width=0.55\textwidth}
%\vspace{-1cm}
\caption{Relative flux of \c\ phased with the orbital period and time of transit of companion b. The dotted line shows the 1-$\sigma$ uncertainty in the transit ephemeris. The thick line shows the expected transit duration and depth for the nominal ephemeris. Transits are ruled out to a depth of $\sim$0.4\%.}
\label{photomc}
\end{center}
\end{figure}

The \c\,b planet has a transit probability of 12\%. Nine series of observations were carried out on 15 and 30 August 2012 and in May to July 2013 with the Oversky, ROTAT, BROA and Crow facilities. BROA uses a 20cm Celestron C8 with a KAF402XME camera and an R$_c$ filter. The Crow observatory uses a 30cm aperture Meade LX200, an I$_c$ filter, and an  SBIG ST8XME camera. Combining the 47 hours of observations of this target (online Table 8), we can exclude that a transit occurred between -9 and +6 hours around the expected times for the transit center.  Figure \ref{photomc} shows the merged photometric light curve obtained at the expected time of transit. The r.m.s. of the measurements is 2 mmag. A transit with 0.4\% depth would have been detected, while a depth of $\sim1$\% is expected for a Jupiter-sized planet passing accross a solar-like disk. 

%The duration of first both series is 5.2 and 6.5 hours, respectively, and their standard deviation of the normalized flux is 1.3\% over the expected transit window. The third sequence has a better accuracy %and is shown on Fig. \ref{photomc} 
%with 0.38\% scatter of the relative flux for a sequence of 5.4 hours. No transit was detected in these data. %Figure \ref{photomc}  shows the data and expected transit duration and ephemeris at the time of observations. 

\begin{table} 
\caption{Orbital and physical parameters for the planet orbiting \c.  }
\label{TablePlanets}
\centering
\begin{tabular}{l l c }
\hline\hline
\multicolumn{1}{l}{\bf Parameter}&
& \bf \c\,b  \\
\hline
$P$ & [days]  							& 5.0505$\pm$0.0015 \\
$T$ & [JD-2400000] 						&56164.275$\pm$0.028 \\
$T_{transit}$& [JD-2400000] 				&56165.565$\pm$0.023\\
  $e$ &            	  						&0.024$\pm$0.014 \\
$\omega$ & [deg]    	 					&276$\pm$64\\
$K$ & [m s$^{-1}$] 	   					&130.9$\pm$1.7 \\
$m_2 \sin{i}$ & [M$_{\mathrm{Jup}}$]  		&1.068$\pm$0.038 \\
$a$ & [AU]    							&0.057$\pm$0.0009 \\
\hline					
$\gamma_1$ & [km s$^{-1}$] 	   			&-9.452$\pm$0.002 \\ 
linear     	&	[m/s/yr  ] 					&-431$\pm$ 13\\
quadratic  &	[m/s/yr$^2$]				&-1032$\pm$109\\
$N_{\mathrm{meas}}$ & 					&27\\
Span & [days]         					&145.8 \\
$\sigma$ (O-C) & [m s$^{-1}$] 	 			&5.97\\
$\chi^2_{red}$& 						&3.1\\
\hline
\label{solc}
\end{tabular}
\end{table}

\section{Discussion}
\label{disc}
The stars \a, \b, and \c\, host multiple-planet systems with at least two companions. The inner companion is a hot/warm-Jupiter-like planet and the outer companion is more massive. Each planet in the three systems thus represents one of the two main populations of giant planets:  hot/warm Jupiters, and distant giant planets.  Planets \a\,c and \b\,c are located inside the ice line of their respective host star. For the system around \c, the outer companion is not yet characterized, since only part of the orbit is observed, although we were able to estimate that its mass is larger than 2.5 \Mjup. Because the ratio of periods (inner/outer) is lower than 0.2, the three systems can be considered as  stable hierarchical systems. Their dynamical evolution is not subject to mean-motion resonance interactions but rather to secular perturbations \citep{beauge12r}.

Their configurations are not common, since most multiple systems are composed of low-mass planets. Other hot/warm Jupiter planets (mass greater than 0.5 \Mjup\ and period shorter than 20 days) with (at least) one known outer companion are: 55 Cnc b, HAT-P-13 b, HAT-P-17 b, HD187123 b, HD 217107 b, HD 38529 b,  HIP 14810 b, and $\upsilon$ And b. All of these systems except  HIP 14810 have outer planets more massive than the inner one, like the systems around \a, \b\ and \c\  (the mass of \c\,c is still undetermined, but it is larger than the mass of \c\,b). 

 Including all detection technics, there are currently 146 known multiple-planet systems\footnote{exoplanet.eu}, of which 92 have (minimum) mass measurements of all companions through radial-velocity measurements.
%Not counting the Kepler systems where the planet mass regime is estimated by the transit timing variations (only the maximum mass allowing system stability is constrained), there are 92 multiple-planet systems\footnote{exoplanet.eu}. 
Ten of these systems have an inner planet of the hot/warm-Jupiter category (with a period shorter than 20 days), which represents 11\% of them. This is compatible with the analysis of \citet{steffen12} from the Kepler-object-of-interest sample \citep{batalha}, who have shown that planets with periods larger than 6.3 days or with a radius smaller than 0.6 \Rjup\  have a $\gtrsim$10\% probability to have another transiting or TTV-planet in the system. %, compared to shorter period planets or more massive ones.

We cannot, however, derive reliable occurrence rates of hot/warm Jupiters in systems with respect to single hot-Jupiters from current observations, because the sample of multiple systems is heterogeneous and detection limits vary from one system to the other. In addition, the detection bias of radial-velocity and transit techniques favours the detection of short-period Jupiters, but severely limits the detection of outer planets: the latter requires both long-term observations and an accuracy of a few \ms\  for the decreasing amplitude; in addition, the transit probability decreases with increasing semi-major axis. 

Multiple systems composed of giant planets can serve as important constraints for migration models. The formation of hot Jupiters is not thought to occur {\it in situ}, but at large distances from their central star, followed by orbital evolution due to several types of interactions in the system: interaction with the proto-planetary disks (type II migration), dynamical interactions with other planets in the system, or Kozai migration. In early papers after the discovery of 51 Peg b, the formation of a hot-Jupiter planet was shown to result from planet-planet scattering followed by tidal circularization of the inner planet \citep{rasio96}, for certain initial conditions. The formation of giant planets with a massive inclined body at large distances can also result in a decrease of the physical separation between the planet and its host star, until the orbit is circularized by tides \citep{kozai62,wu07}. The detection and characterization of systems with a hot/warm Jupiter and another outer body in the system therefore gives practical examples of the potential outcome of these mechanisms.  To determine whether  type II migration is the most common phenomenon for the formation of hot Jupiters, as was recently discussed by \citet{rice12}, it would be necessary to have precise observational constraints on all single hot-Jupiter systems, to determine what types of  massive outer bodies are allowed by the data. A full picture of planetary system architecture is required to understand the origins of the inner-planet migration. Interestingly, the inner planets \a\, b, \b\, b, and \c\, b found in the present work do not have periods corresponding to the pile-up of hot Jupiters at a period of about three days \citep[e.g.,]{udry03}, but longer periods of 5 to 20 days. This may be a sign that their current distance to their star originates in an evolution induced by other mechanims than friction with the protoplanetary disks, such as planet-planet interactions (for \a\, b and \b\, b) or planet-star interactions (for \c\, b, whose system may have a planetary or stellar outer companion).

The occurrence of massive planets at about 2 AU, such as \a\,c (mass greater than 5 \Mjup) may also be an interesting constraint for the accretion model  with migration, beause it was shown by \citet{alexander12} that the efficiency of accretion across the gap has a strong influence on the existence of such planets. Moreover, the planet \b\,c with about 2 \Mjup\ at 0.8 AU may correspond to the predicted pile-up of planets caused by photoevaporative clearing of the disk, as predicted in \citet{alexander12} and observationally shown by \citet{wright09}.

The relatively large semi-major axis of \a\,b and \b\,b  fits the planet-planet scattering models with two initial planets better than those with three or four planets, if the orbital evolution of these systems is due to dynamical interactions between the planets and in the context of the simulations performed and discussed by \citet{beauge12}. 

%Finally, we note the \c\,b is a hot Jupiter around a metal-rich star. It strengthens the well-established fact that metal-rich stellar hosts show a larger occurrence rate of giant planets than metal-poor stars \citep{santos04, fischer05}. 

Our study confirmed the interest in extending the observational constraints of the exoplanet sample, in particular by following-up stars with known planets over a  time scale of several years, at a precision of a few \ms. This allows one, in the long term, to assemble a more general picture of the mass distribution in the system after its dynamical evolution.

\acknowledgements{We gratefully acknowledge the Programme National de Plan\'etologie (telescope time attribution and financial support), the Swiss National Foundation, and the Agence Nationale de la Recherche (grant ANR-08-JCJC-0102-01) for their support. R.F.D. is supported by CNES. We warmly thank the OHP staff for their great care in optimising the observations. N.C.S., I.B. and A.S. acknowledge the support of the European Research Council/European Community under the FP7 through Starting Grant agreement number 239953. N.C.S., I.B. and A.S. also acknowledge the support from Fundacao para a Ci\^encia e a Tecnologia (FCT) through program Ci\^encia\,2007 funded by FCT/MCTES (Portugal) and POPH/FSE (EC), and in the form of grants reference PTDC/CTE-AST/66643/2006, PTDC/CTE-AST/098528/2008, and SFRH/BPD/81084/2011. }

\onltab{5}{
\begin{table}[b]
\tiny
\centering
  \caption{Radial-velocity measurements obtained with SOPHIE of \a.}
  \label{rva}
\begin{tabular}{lccc}
\hline
JD-2,400,000.  &  Radial Vel. & Uncertainty & Bis Span\\
         & [km s$^{-1}$]   &  [km s$^{-1}$] &  [km s$^{-1}$] \\
\hline
54767.50725  &  15.7709  &  0.004096  &  0.0485  \\
54809.37187  &  15.7398  &  0.003576  &  0.0588  \\
55434.64430  &  15.8987  &  0.005186  &  0.0507  \\
55434.64678  &  15.8843  &  0.002708  &  0.0473  \\
55557.31431  &  15.9048  &  0.005970  &  0.0632  \\
55577.30254  &  15.8711  &  0.002559  &  0.0435  \\
55578.27211  &  15.8606  &  0.002622  &  0.0432  \\
55579.33908  &  15.8671  &  0.003680  &  0.0587  \\
55580.27264  &  15.8286  &  0.003047  &  0.0383  \\
55583.33764  &  15.7993  &  0.005814  &  0.0338  \\
55587.30206  &  15.7681  &  0.005788  &  0.0577  \\
55587.30370  &  15.7798  &  0.005781  &  0.0810  \\
55597.30251  &  15.8581  &  0.005345  &  0.0425  \\
55597.30473  &  15.8604  &  0.003056  &  0.0478  \\
55599.30208  &  15.8431  &  0.003126  &  0.0552  \\
55620.32595  &  15.8071  &  0.005640  &  0.0368  \\
55627.28875  &  15.7511  &  0.003263  &  0.0445  \\
55630.27425  &  15.8136  &  0.004070  &  0.0385  \\
55631.29515  &  15.8149  &  0.003895  &  0.0362  \\
55638.31487  &  15.8406  &  0.006438  &  0.0390  \\
55640.28324  &  15.7973  &  0.005869  &  0.0500  \\
55646.28013  &  15.7445  &  0.003989  &  0.0450  \\
55646.28340  &  15.7362  &  0.004105  &  0.0452  \\
55646.28587  &  15.7423  &  0.004060  &  0.0588  \\
55647.28173  &  15.7548  &  0.006255  &  0.0732  \\
55647.28423  &  15.7494  &  0.005590  &  0.0468  \\
55649.28213  &  15.8007  &  0.004157  &  0.0610  \\
55649.28510  &  15.7988  &  0.004132  &  0.0467  \\
55651.27797  &  15.8268  &  0.007974  &  0.0642  \\
55651.28227  &  15.8169  &  0.005876  &  0.0510  \\
55659.28847  &  15.7956  &  0.005162  &  0.0580  \\
55663.29311  &  15.7276  &  0.005844  &  0.0500  \\
55665.31366  &  15.7451  &  0.009224  &  0.0592  \\
55668.30748  &  15.7981  &  0.007173  &  0.0258  \\
55671.32016  &  15.8286  &  0.076182  &  0.5858  \\
55672.30442  &  15.8348  &  0.008317  &  0.0490  \\
55672.30767  &  15.8298  &  0.007144  &  0.0210  \\
55673.31200  &  15.8304  &  0.009507  &  0.0373  \\
55679.32709  &  15.7631  &  0.007806  &  0.0433  \\
55683.32105  &  15.7058  &  0.010255  &  0.0627  \\
55686.31406  &  15.7582  &  0.007617  &  0.0703  \\
\hline
55785.59289  &  15.8281  &  0.005904  &  0.0387  \\
55788.63021  &  15.8639  &  0.003961  &  0.0388  \\
55795.61180  &  15.7901  &  0.005788  &  0.0538  \\
55798.65008  &  15.7590  &  0.005889  &  0.0440  \\
55799.64490  &  15.7546  &  0.005931  &  0.0663  \\
55805.65203  &  15.8464  &  0.005956  &  0.0498  \\
55813.57298  &  15.8410  &  0.005816  &  0.0472  \\
55814.61203  &  15.8185  &  0.005845  &  0.0395  \\
55814.61446  &  15.8190  &  0.003929  &  0.0650  \\
55818.47529  &  15.7808  &  0.005786  &  0.0792  \\
55827.53043  &  15.8895  &  0.005861  &  0.0372  \\
55835.55372  &  15.7972  &  0.003846  &  0.0517  \\
55842.55815  &  15.8331  &  0.005923  &  0.0245  \\
55852.50272  &  15.8490  &  0.004825  &  0.0767  \\
55856.63494  &  15.8036  &  0.004397  &  0.0355  \\
55866.44310  &  15.9136  &  0.005931  &  0.0727  \\
55878.38808  &  15.8155  &  0.004775  &  0.0468  \\
55882.45955  &  15.8818  &  0.005872  &  0.0452  \\
55904.31050  &  15.9184  &  0.004198  &  0.0423  \\
55916.39987  &  15.8392  &  0.005975  &  0.0713  \\
55926.29334  &  15.9501  &  0.006051  &  0.0603  \\
55938.31974  &  15.8883  &  0.005845  &  0.0498  \\
55957.26549  &  15.8872  &  0.005816  &  0.0403  \\
55964.33337  &  15.9702  &  0.006063  &  0.0313  \\
55976.25564  &  15.8803  &  0.006028  &  0.0392  \\
55997.29594  &  15.9218  &  0.006099  &  0.0733  \\
56017.28944  &  15.9437  &  0.003866  &  0.0388  \\
56149.62463  &  15.9181  &  0.003168  &  0.0435  \\
56157.62640  &  16.0270  &  0.003022  &  0.0420  \\
56179.50970  &  16.0134  &  0.006138  &  0.0522  \\
56188.58121  &  15.9200  &  0.003302  &  0.0452  \\
56217.39032  &  16.0106  &  0.005880  &  0.0365  \\
56234.46018  &  16.0210  &  0.003412  &  0.0427  \\
56263.43382  &  15.9038  &  0.005837  &  0.0502  \\
%\hline
%56297.26618  &  15.9801  &  0.005762  &  0.0430  \\
\hline
\end{tabular}
\end{table}
}

\onltab{6}{
\begin{table}[b]
\tiny
\centering
  \caption{Radial-velocity measurements obtained with SOPHIE of \b.}
  \label{rvb}
\begin{tabular}{lccc}
\hline
JD-2,400,000.  &  Radial Vel. & Uncertainty& Bis Span \\
         & [km s$^{-1}$]   &  [km s$^{-1}$] &  [km s$^{-1}$] \\
\hline
55638.66550  &  -22.7778  &  0.005466  &  -0.0147  \\
55640.64998  &  -22.7183  &  0.005188  &  0.0188  \\
55641.65382  &  -22.6864  &  0.004397  &  -0.0042  \\
55660.64431  &  -22.7676  &  0.004966  &  0.0253  \\
55662.66170  &  -22.7938  &  0.005066  &  0.0003  \\
55667.63040  &  -22.6122  &  0.044468  &  -0.0573  \\
55668.57948  &  -22.6022  &  0.007755  &  0.0077  \\
55669.53776  &  -22.6204  &  0.005075  &  0.0213  \\
55670.56762  &  -22.6461  &  0.004887  &  -0.0002  \\
55671.63092  &  -22.6844  &  0.005146  &  0.0037  \\
55672.59817  &  -22.7358  &  0.004716  &  -0.0030  \\
55673.60054  &  -22.7637  &  0.005246  &  -0.0122  \\
55675.59222  &  -22.7896  &  0.005016  &  -0.0085  \\
55676.63612  &  -22.7635  &  0.005298  &  -0.0095  \\
55677.57695  &  -22.7244  &  0.004942  &  0.0228  \\
55679.53866  &  -22.6354  &  0.005153  &  0.0048  \\
55681.50791  &  -22.5982  &  0.005064  &  0.0115  \\
55682.62087  &  -22.5897  &  0.005188  &  0.0063  \\
55683.51246  &  -22.6375  &  0.004798  &  0.0013  \\
55697.55415  &  -22.6494  &  0.008064  &  0.0235  \\
55702.44127  &  -22.7242  &  0.004866  &  -0.0073  \\
55706.45624  &  -22.5506  &  0.005189  &  -0.0127  \\
\hline
55745.38996  &  -22.5234  &  0.005939  &  -0.0060  \\
55762.35288  &  -22.6625  &  0.005067  &  0.0400  \\
55786.40889  &  -22.6685  &  0.005246  &  0.0257  \\
55793.32333  &  -22.5896  &  0.005282  &  0.0117  \\
55799.34877  &  -22.7004  &  0.005336  &  0.0030  \\
55807.29972  &  -22.5930  &  0.005261  &  0.0042  \\
55809.35545  &  -22.6076  &  0.005271  &  0.0168  \\
55812.29685  &  -22.7347  &  0.005210  &  -0.0263  \\
55817.28854  &  -22.6841  &  0.005230  &  0.0162  \\
55818.29233  &  -22.6433  &  0.005236  &  0.0352  \\
55821.28792  &  -22.6100  &  0.005261  &  -0.0282  \\
55823.29087  &  -22.6724  &  0.005409  &  0.0058  \\
55824.29376  &  -22.7166  &  0.005351  &  0.0087  \\
55825.28511  &  -22.7640  &  0.005414  &  0.0105  \\
55827.28447  &  -22.7866  &  0.005347  &  -0.0113  \\
55840.28498  &  -22.7908  &  0.005293  &  0.0168  \\
55841.26794  &  -22.7599  &  0.005914  &  0.0302  \\
55850.25186  &  -22.7810  &  0.005410  &  0.0115  \\
55878.23149  &  -22.7905  &  0.006266  &  -0.0050  \\
56016.65959  &  -22.7076  &  0.005434  &  0.0095  \\
56054.51930  &  -22.7463  &  0.005390  &  0.0013  \\
56072.54677  &  -22.5910  &  0.004999  &  -0.0040  \\
56087.53884  &  -22.6613  &  0.005896  &  -0.0017  \\
56107.44685  &  -22.6971  &  0.004085  &  0.0097  \\
56140.35575  &  -22.7320  &  0.004172  &  0.0212  \\
56150.32715  &  -22.6099  &  0.005339  &  0.0163  \\
56164.34751  &  -22.6744  &  0.005443  &  -0.0110  \\
56179.35399  &  -22.7217  &  0.005514  &  -0.0055  \\
56217.25185  &  -22.6838  &  0.005286  &  0.0157  \\
\hline
\end{tabular}
\end{table}
}

\onltab{7}{
\begin{table}[b]
\tiny
\centering
  \caption{Radial-velocity measurements obtained with SOPHIE of \c.}
  \label{rvc}
\begin{tabular}{lccc}
\hline
JD-2,400,000.  &  Radial Vel. & Uncertainty& Bis Span \\
         & [km s$^{-1}$]   &  [km s$^{-1}$] &  [km s$^{-1}$] \\
\hline
56098.51190  &  -9.2761  &  0.003419  &  -0.0425  \\
56106.44881  &  -9.5369  &  0.003159  &  -0.0258  \\
56113.44030  &  -9.2840  &  0.003516  &  -0.0310  \\
56116.38176  &  -9.5416  &  0.003244  &  -0.0347  \\
56117.39637  &  -9.4548  &  0.003378  &  -0.0313  \\
56121.47155  &  -9.5586  &  0.003389  &  -0.0390  \\
56122.42013  &  -9.4619  &  0.003392  &  -0.0237  \\
56124.35294  &  -9.3156  &  0.003435  &  -0.0357  \\
56137.46212  &  -9.4781  &  0.003403  &  -0.0310  \\
56138.40492  &  -9.3365  &  0.003384  &  -0.0297  \\
56140.41455  &  -9.4384  &  0.003634  &  -0.0072  \\
56141.41877  &  -9.5549  &  0.003171  &  -0.0285  \\
56150.33240  &  -9.4283  &  0.003407  &  -0.0338  \\
56151.32785  &  -9.5635  &  0.003699  &  -0.0308  \\
56152.34043  &  -9.5338  &  0.003484  &  -0.0298  \\
56155.38439  &  -9.4415  &  0.003382  &  -0.0367  \\
56162.31653  &  -9.5582  &  0.003383  &  -0.0282  \\
56164.42055  &  -9.3237  &  0.003534  &  -0.0285  \\
56172.34330  &  -9.5742  &  0.003447  &  -0.0300  \\
56179.36152  &  -9.3549  &  0.003465  &  -0.0273  \\
56191.32642  &  -9.5652  &  0.003415  &  -0.0145  \\
56219.26892  &  -9.4456  &  0.003458  &  -0.0142  \\
56225.25973  &  -9.4413  &  0.003421  &  -0.0420  \\
56230.30687  &  -9.4572  &  0.003364  &  -0.0302  \\
56231.23900  &  -9.5736  &  0.003082  &  -0.0280  \\
56237.24078  &  -9.7096  &  0.003514  &  -0.0453  \\
56244.29209  &  -9.5361  &  0.003422  &  -0.0347  \\
\hline
\end{tabular}
\end{table}
}

\bibliographystyle{aa}
\bibliography{s2013}

\end{document}